# Spin-Torque Driven Magnetization Dynamics: Micromagnetic Modelling

D.V. Berkov[1], J. Miltat[2]

[1] *Innovent Technology Development, Pruessingstr. 27B, D-07745 Jena, Germany*
[2] *Laboratoire de Physique des Solides, Univ. Paris-Sud & CNRS, 91405 Orsay, France*

## Abstract

In this paper we present an overview of recent progress made in the understanding of the spin-torque induced magnetization dynamics in nanodevices using mesoscopic micromagnetic simulations. We first specify how a spin-torque term may be added to the usual Landau-Lifshitz-Gilbert equation of magnetization motion and detail its physical meaning. After a brief description of spin-torque driven dynamics in the *macrospin approximation*, we discuss the validity of this approximation for various experimentally relevant geometries. Next, we perform a detailed comparison between accurate experimental data obtained from *nanopillar* devices and corresponding numerical modelling. We show that, on the one hand, many qualitatively important features of the observed magnetization dynamics (e.g., non-linear frequency shift and frequency jumps with increasing current) can be satisfactory explained by sophisticated micromagnetic models, but on the other hand, understanding of these experiments is still far from being complete. We proceed with the numerical analysis of *point-contact experiments*, where an even more complicated magnetization dynamics is observed. Simulations reveal that such a rich behaviour is due to the formation of several strongly non-linear oscillation modes. In the last part of the paper we emphasize the importance of sample characterization and conclude with some important remarks concerning the relation between micromagnetic modelling and real experiments.

.

# Content:

# I. Introduction

An intuitive approach to micromagnetics might consider a ferromagnet as an assembly of localized dipoles governed by Heisenberg exchange interactions, dipole-dipole interactions as well as anisotropy, a characteristic of magnetic bodies tightly linked to properties of the orbital moment. The distribution of magnetization within tiny magnetic islands grown on favourable substrates in ultra-high vacuum (e.g. [1]) may be studied in such a way [2]. Classical micromagnetics [3, 4, 5] treats the magnetization in the continuum limit; assuming a constant saturation magnetization for the operating temperature, the magnetization distribution within a ferro- or ferrimagnetic body is thus becoming a vector-field $\mathbf{M}(\mathbf{r}) = M_S \mathbf{m}(\mathbf{r})$, where $\mathbf{m}(\mathbf{r})$ may only reside on the unit sphere $|\mathbf{m}(\mathbf{r})| = 1$. Dipole-dipole interactions are replaced by magnetostatics, a formal equivalent to electrostatics with the necessary condition that magnetic "charges" sum-up to zero, akin a simple dipole. Magnetic "charges" arise either from divergences of the magnetization vector-field within the volume of the ferromagnetic body ($\nabla \mathbf{m}(\mathbf{r}) \neq 0$), or, assuming so-called "free boundary" conditions to apply, as soon as the magnetization vector fails to be parallel to a free surface or an interface ($(\mathbf{m} \cdot \mathbf{n}) \neq 0$, where $\mathbf{n}$ is the outwards normal to the surface or interface). Heisenberg-type exchange interactions are taken in the continuous limit and assumed isotropic. Therefore, a single parameter describes the stiffness of the magnetization vector field to distortions whether bending, splay or twist. Anisotropy is described on a purely phenomenological basis. Lastly, starting from a magnetization distribution at equilibrium, the magnetization may be set into motion under the action of an external field, often called the applied field, or a spin-polarized current [6, 7]. Both exert a torque on the magnetization.

An energy density is associated to each type of interaction within the magnetic body, spin-torque omitted. A magnetization distribution is thus the result of conflicting requirements. Exchange interactions promote uniform magnetization distributions. On the other hand, imposing the condition $(\mathbf{m} \cdot \mathbf{n}) = 0$ along the boundaries of a magnetic body necessarily leads to non-uniform distributions. Consider for instance a flat cylindrical micron size platelet: imposing the condition $(\mathbf{m} \cdot \mathbf{n}) = 0$ states that the magnetization should remain in-plane and tangential to the rim of the platelet. Imposing the condition $\nabla \mathbf{m}(\mathbf{r}) = 0$ means that, away from the rim, the magnetization should remain orthogonal to any radius drawn from the cylinder axis to the rim. When approaching the cylinder axis, however, such a circular magnetization distribution, called a vortex, leads to increasing exchange interactions, and, soon to a divergence of the exchange energy. Because of the constraint $|\mathbf{m}(\mathbf{r})| = 1$, the magnetization needs to pop-up out of the plane. Due to symmetry, the magnetization direction along the core of the vortex may only be perpendicular to the plane of the platelet, up or down. The transition from in-plane to out-of-plane magnetization orientation takes place over distances not exceeding a few nanometers (see e.g. [8] for a short introduction to vortices). More generally, in the absence of any spin-polarized current, a magnetization distribution reaches equilibrium when the energy reaches its minimal value. When not submitted to an applied field or spin-polarized current, any magnetization distribution is energetically equivalent to the distribution obtained through the transformation $\mathbf{m}(\mathbf{r}) \rightarrow -\mathbf{m}(\mathbf{r})$. Additional degeneracy may arise from the geometrical symmetries of micron or sub-micron size magnetic elements.

Numerical micromagnetics as a tool proves adequate for a fine description of magnetization distributions at remanence within finite magnetic bodies or their transformation under the action of a steady field (see [9] for recent reviews of numerical techniques). Just to quote a single example, magnetization distributions observed (MFM and X-PEEM experiments) in thick, facetted, single crystalline Fe islands grown on a Mo (110) surface [10] although intriguing at first sight appear to closely fit micromagnetic simulations taking due account of the shape features of these (nearly) perfect crystals. Although not fully revealed by experiments, the internal wall structure is anticipated to still prove complex. In the following we do assume that fundamental material parameters such as the saturation magnetization $M_S$ at the working temperature, anisotropy constants such as $K_{un}$ (characteristic of a uniaxial anisotropy), $K_1$ and $K_2$ (characteristic of a cubic anisotropy) etc, and the exchange constant $A$, are known with an accuracy sufficient not to impair simulation results. Such quantities may, in principle, be known independently via, e.g., SQUID magnetometry, torque magnetometry and spin waves analysis, respectively.

The primary aim of this "Perspective" is the evaluation of the ability of micromagnetics to describe magnetization dynamics under spin-torque excitation. As a prerequisite, however, it seems worthwhile examining whether numerical micromagnetics remains accurately predictive in the field-driven *dynamical* regime. On top of material parameters listed above, one now also needs, as shown below in Sec. II, to have at hand a fair evaluation of the gyromagnetic ratio $\gamma$, as well as the damping constant $\alpha$. Rare are the experiments where all of the quantities that enter the Landau-Lifshitz or Landau-Lifshitz-Gilbert equation of magnetization motion (see Sec. II below) are known accurately. For instance, in their study of modes in the vortex state, Buess et al. claim a remarkable agreement between experimental data and micromagnetic simulations, with an experimental accuracy better than 10% [11]. However, the authors' choice of a free electron value for $\gamma$ is most likely inappropriate. Arguably, though, their choice of a rather high saturation magnetization for Permalloy at room temperature may well compensate for the likely ≈ 5% or more error in the gyromagnetic ratio. An interesting outcome of this study is that when the magnetization is 'tipped' away from its original orientation and then allowed to relax towards equilibrium in the absence of any applied field, then one obtains the same value of the damping parameter, irrespective of the active mode ($\alpha \approx 0.008$ for the samples studied). Similarly, Novosad et al. [12] found an excellent agreement between the vortex gyration frequency and numerical simulations, also in Permalloy disks (under gyration we mean here vortex motion along an essentially circular closed orbit). Here experimental data represent an average over a large number of similar magnetic elements. Simulations do rely on a rather depressed (very thin samples) saturation magnetization measured independently and on a free electron value of the gyromagnetic ratio as well as a "standard" value of the damping parameter in Permalloy ($\alpha = 0.01$). If eigenmode frequencies most often amount to a few GHz, vortex gyration frequencies depend, for a given thickness, on the disk diameter and do not exceed a few hundreds of MHz. Eigenmodes analysis relies on small amplitude motion of the magnetization whereas vortex gyration implies larger amplitude motion. Work quoted above altogether displays a very satisfactory agreement between measured and computed frequencies (see also [13, 14] for recent studies of vortex motion and vortex core reversal under field or spin-polarized current).

This, however, is not always the case: in their study of localized spin-wave modes in micron-wide magnetic wires, Park et al. [15] do reach good agreement between experiments and simulations for fields applied in the plane of and normal to the wire edge in excess of ≈ 100 Oe. Below that value, experimental frequencies prove significantly lower than predicted by simulations. Similarly, a systematic discrepancy between experiments and simulations has been found in the study of eigenmodes in Co rings [16]. In this particular case, however, computed frequencies prove lower than experimental values. Such facts probably need to be borne in mind when analyzing the outcome of simulations picturing spin-transfer torque driven magnetization dynamics.

Lastly, since sustained precession under the action of a steady spin-polarized current may imply large angle precession, it would have been extremely useful to compare experimental data dealing with ballistic switching (also implying large angle precession) [e.g. 17, 18] and detailed micromagnetic simulations. Rather unfortunately in a sense, magnetization dynamics in the single-spin limit with ad-hoc parameters proved sufficient for a very stunning description of phase coherent ballistic switching in GMR or TMR micron size elements. Nevertheless, it altogether appears that whereas micromagnetics proves tightly predictive when describing the static behaviour of magnetic elements, it performs less satisfactorily when attempting a quantitative analysis of magnetization dynamics.

Let us now move to the core of this paper. Because sustained precession in CPP-GMR nanopillars and point contacts leads to precise observables, namely a well defined current- and field-dependent frequency, and an associated power spectral density, the present perspective is limited to spin-transfer torque induced magnetization dynamics in these structures, leaving aside spin pressure effects on walls in magnetic nanowires. It is organized as follows. Sec. II recalls which are the minimal modifications to be brought to the equation of magnetization motion in order to take into account spin-torque effects. In Sec. III, we attempt to analyse reasons why a single spin approximation fails to provide an acceptable picture of observed phenomena in the structures considered. The following section illustrates how different working hypotheses may lead to markedly different micromagnetics simulation results when dealing with nano-pillar geometries. Peculiarities of the point contact geometry are analyzed in Sec. V. Finally we discuss in Sec. VI the crucial importance of sample characterization for a meaningful numerical analysis of spin-torque experiments, before reaching concluding remarks.

## II. Including spin-torque effects into magnetization dynamics

Almost all of micromagnetic simulations involving magnetization dynamics rely on the Landau-Lifshitz-Gilbert equation of magnetization motion, namely (if using SI units)

$$\frac{d\mathbf{M}(\mathbf{r},t)}{dt} = -\gamma_0 \left[ \mathbf{M}(\mathbf{r},t) \times \mathbf{H}_{eff}(\mathbf{r},t) \right] + \alpha \left[ \mathbf{M}(\mathbf{r},t) \times \frac{d\mathbf{M}(\mathbf{r},t)}{dt} \right]$$

$$\mathbf{H}_{eff}(\mathbf{r},t) = -\frac{1}{\mu_0} \frac{\delta \varepsilon}{\delta \mathbf{M}},$$

(1)

where $\mathbf{M}(\mathbf{r},t)$ and $\mathbf{H}_{eff}(\mathbf{r},t)$ are the magnetization and effective field, respectively. Both are functions of space and time. $\varepsilon$ denotes the energy density functional, $\gamma_0 = \mu_0 \cdot (g\mu_B/\hbar)$ ($\gamma_0 \cong 2.211 \cdot 10^5$ mA$^{-1}$s$^{-1}$ for a free electron), $\mu_B$ is the Bohr magneton and $\alpha$ the damping parameter. The effective field is the sum of the applied, anisotropy and demagnetizing fields, supplemented by field components arising from exchange interactions. Eq. (1) means that precession around the local effective field is the fundamental magnetization motion. Note, however, that the effective field moves together with the magnetization, and, thus, the simple idea of a precession around a field with *fixed* direction may prove extremely misleading. Damping is required in order to align the magnetization along the acting field: the Gilbert form used in (1) is consistent with Rayleigh-type dissipation. As noticed numerous times before, Eq. (1) is strictly equal to:

$$\frac{d\mathbf{M}(\mathbf{r},t)}{dt} = -\frac{\gamma_0}{1+\alpha^2} \left[ \mathbf{M}(\mathbf{r},t) \times \mathbf{H}_{eff}(\mathbf{r},t) + \frac{\alpha}{M_S} \mathbf{M}(\mathbf{r},t) \times [\mathbf{M}(\mathbf{r},t) \times \mathbf{H}_{eff}(\mathbf{r},t)] \right],$$

(2)

thus recovering the initial damping formulation of the Landau-Lifshitz equation at the expense of a (minor since $\alpha \ll 1$, usually) renormalization of both the gyromagnetic ratio and the damping parameter. Eq. (2) may also been written as:

$$\frac{d\mathbf{M}(\mathbf{r},t)}{dt} = -\frac{\gamma_0}{1+\alpha^2}\Big[\mathbf{M}(\mathbf{r},t)\times\big(\mathbf{H}_{\text{eff}}(\mathbf{r},t)+\mathbf{H}_{\text{damp}}(\mathbf{r},t)\big)\Big]$$
$$\mathbf{H}_{\text{eff}}(\mathbf{r},t) = -\frac{1}{\mu_0}\frac{\delta\varepsilon}{\delta\mathbf{M}}; \quad \mathbf{H}_{\text{damp}}(\mathbf{r},t) = \frac{\alpha}{M_S}\big[\mathbf{M}(\mathbf{r},t)\times\mathbf{H}_{\text{eff}}(\mathbf{r},t)\big] \quad (3)$$

As noticed by N. Smith [19], Eq. (3) treats on an equal footing two field terms: the effective field that is conservative (it derives from an energy density functional) and a field that, per definition, is non conservative (energy is transferred to an external bath).

In the presence of an electric current, an additional torque may act on the magnetization within a thin ferromagnetic layer, arising primarily from the transmission and reflection of incoming electrons with moments at arbitrary angles to the magnetization [6, 20, 21, 22]. As a net result, reflected and transmitted spin currents have virtually no component transverse to the magnetization. In other words, this pure ballistic effect leads to a close to complete absorption of the transverse spin current, itself the source of the spin-transfer torque (see [23] for a review of concepts). In a CPP-GMR stack, the electrons acquire spin polarization either because they first cross the pinned (or hard) layer of the stack, or because they get reflected from the latter. Let $\mathbf{p}$ and $\mathbf{m} = \mathbf{M}/M_S$ be the unit vectors along the magnetization of the pinned and soft layers of the stack, respectively. For these reasons, the spin transfer torque is proportional to the sine of the angle between $\mathbf{p}$ and $\mathbf{m}$, or in vector notation to $[\mathbf{m}\times[\mathbf{m}\times\mathbf{p}]]$. The spin transfer torque is also proportional to the quantum of angular momentum carried by one electron times the density of carriers per unit time weighted by the electron polarization $P$, i.e. $\propto P\frac{\hbar}{2}\frac{J}{|e|}[\mathbf{m}\times[\mathbf{m}\times\mathbf{p}]]$, where $J$ is the current density and $e$ the electron charge ($e<0$).

In CPP-GMR stacks, the electron polarization at the interface between the normal metal spacer and the ferromagnetic free layer is affected by the spin-dependent transport characteristics of the whole stack, due to (*i*) spin relaxation within the bulk of the layers or at interfaces, and (*ii*) spin accumulation [24]. Slonczewski proposed in 2002 an elegant approach to this problem where spin-flip scattering is not allowed within the spacer layer [25]. This "full acceptance" model may be developed into a simple circuit theory that leads to immediately usable expressions for both the CPP-GMR and spin transfer torque in the particular case of "symmetrical" spin valves, i.e. stacks where the two FM layers are made of the same material with equal thicknesses, and leads are also made of a unique material and have equal lengths. For these reasons, Slonczewski's vintage'2002 model soon became popular, although it needs to be realized that fulfilling the condition of two identical ferromagnetic layers with an identical environment does not easily fit with the necessity to pin or harden the magnetization of one of the ferromagnetic layers in order for a CPP-GMR device to be functional. According to this Slonczewski's model, the spin transfer torque may be written as:

$$\frac{d\mathbf{M}}{dt} = -\gamma_0\left(\frac{\hbar}{2}\frac{1}{\mu_0 M_S^2}\frac{1}{d}\frac{J}{e}\right)\cdot\frac{P^{\text{Sloncz}}/2}{\cos^2(\vartheta/2)+\frac{1}{1+\chi_a}\sin^2(\vartheta/2)}\cdot[\mathbf{M}\times[\mathbf{M}\times\mathbf{p}]], \quad (4)$$

(still assuming $e < 0$)[1], whereas the GMR response turns out to be proportional to

$$\frac{1-\cos^2(\vartheta/2)}{1+\chi_a \cos^2(\vartheta/2)} \quad (5)$$

In Eqs. (4) and (5), $\vartheta$ is the angle between $\mathbf{m}$ and $\mathbf{p}$ ($\cos\vartheta = (\mathbf{m}\cdot\mathbf{p})$); the asymmetry parameter $\chi_a$ and the polarization $\mathsf{P}^{\text{Sloncz}}$ do both depend on the stack-circuit characteristics and may be expressed as combinations of spin-dependent bulk and interface resistances. The current is negative when electrons first cross the "pinned", then the "free" layer, and positive otherwise. Finally, it ought to be understood that the spin transfer torque has been distributed over the thickness $d$ of the "free" or "soft" layer in Eq. (4). Eqs. (4) and (5) do capture essential ingredients of spin polarized transport in CPP-GMR stacks, namely an asymmetry of the GMR response as well as an asymmetry of the spin torque. More detailed calculations [26] did not uncover significant deviations from Slonczewski's model for materials and thicknesses entering typical CPP-GMR nano-pillars.

Most simulations relying on Slonczewski's or Xiao/Zangwill/Stiles' analysis of spin torque effects in magnetic nanostructures forget about the meaning of the polarization factor and replace $(1/2)\mathsf{P}^{\text{Sloncz}}$ by a simple polarization-like adjustable parameter $\mathsf{P}_{\text{eff}}$. Several additional remarks ought to be made:

(i) cumbersome physical constants may be avoided *via* variable reduction. In SI units, defining reduced variables as $\chi = \frac{\hbar}{2}\frac{1}{\mu_0 M_S^2}\frac{1}{d}\frac{J}{e}\mathsf{P}_{\text{eff}}$, $\tau = \gamma_0 M_S t$, $\mathbf{m} = \mathbf{M}/M_S$, $\mathbf{h} = \mathbf{H}/M_S$, leads to the simple magnetization dynamics equation including spin-torque:

$$\frac{d\mathbf{m}}{d\tau} = -[\mathbf{m}\times\mathbf{h}_{\text{eff}}] - \chi g(\mathbf{m}\cdot\mathbf{p})[\mathbf{m}\times[\mathbf{m}\times\mathbf{p}]] + \alpha\left[\mathbf{m}\times\frac{d\mathbf{m}}{d\tau}\right] \quad (6)$$

$$\mathbf{h}_{\text{eff}} = -\frac{1}{\mu_0 M_S^2}\frac{\delta\varepsilon}{\delta\mathbf{m}}; \quad g(\mathbf{m}\cdot\mathbf{p}) = \left(\cos^2(\vartheta/2) + \frac{1}{1+\chi_a}\sin^2(\vartheta/2)\right)^{-1}$$

Note that both $\chi$ and $\alpha$ are "small" parameters and that $g(\mathbf{m},\mathbf{p})$ does not depart strongly from 1 for common asymmetry parameter values, implying that magnetization trajectories will to a good approximation be determined by the energy landscape.

(ii) As alluded to earlier, the validity of the Gilbert damping term in the presence of spin torque is being debated [27, 19]. Because this perspective is primarily concerned with sustained precession in the nano-pillar and point-contact geometries, the following simple argument shows that a Gilbert (embedding the whole of $d\mathbf{m}/d\tau$) or Landau-Lifshitz formulation restricted to the conservative component of the field, $\mathbf{h}_{\text{eff}}$, may not lead to

$$\frac{d\mathbf{M}}{dt} = -\gamma_{\text{cgs}}\left(\frac{\hbar}{2}\frac{1}{4\pi M_S^2}\frac{1}{d}\frac{J_e}{e}\right)\cdot\frac{\mathsf{P}_{\text{eff}}}{\cos^2(\vartheta/2) + \frac{1}{1+\chi_a}\sin^2(\vartheta/2)}\cdot[\mathbf{M}\times[\mathbf{M}\times\mathbf{p}]],$$

where $\gamma_{\text{cgs}} = g\mu_B^{\text{cgs}}/\hbar^{\text{cgs}} = (10^3/4\pi)\gamma_0$ ($\gamma_{\text{cgs}} \cong 1.76\cdot 10^7$ $(\text{Oe}\cdot\text{s})^{-1}$ for a free electron) and all quantities are expressed in cgs units.

---

[1] For cgs units addicts, Eqn. (4) becomes:

significant differences. Starting with Eq. (6) and taking the cross product of both parts with $d\mathbf{m}/d\tau$ yields:

$$\mathbf{h}_{\text{eff}} \cdot \frac{d\mathbf{m}}{d\tau} + \chi g(\mathbf{m}\cdot\mathbf{p})(\mathbf{m}\times\mathbf{p})\cdot\frac{d\mathbf{m}}{d\tau} - \alpha\left[\frac{d\mathbf{m}}{d\tau}\right]^2 = 0 \qquad (7a)$$

Expression (7a) means that the work of the effective field augmented with the work of the spin torque equivalent field, $\mathbf{h}_{\text{ST}} = \chi\cdot g(\mathbf{m}\cdot\mathbf{p})(\mathbf{m}\times\mathbf{p})$, is balanced by dissipation, at any arbitrary time and at any location within the ferromagnetic body. This is the Gilbert picture. Expression (7a) may also, according to (2), be written as:

$$\{\mathbf{h}_{\text{eff}} + \mathbf{h}_{\text{ST}} + \alpha[\mathbf{m}\times(\mathbf{h}_{\text{eff}} + \mathbf{h}_{\text{ST}})]\} \cdot \frac{d\mathbf{m}}{d\tau} = 0 \qquad (7b),$$

stating that the work of the total field acting on the magnetization, $\mathbf{h}_{\text{tot}} = \mathbf{h}_{\text{eff}} + \mathbf{h}_{\text{ST}}$, is again equilibrated by the work of the total equivalent damping field. Relation (7b) is more in tune with the Landau-Lifshitz approach, which defines the required additional damping field in order to relax the magnetization towards equilibrium without altering the magnitude of the magnetization. In (7b), however, the damping field still arises from both the effective field and the spin-torque equivalent field.

Since the effective field is conservative, for a closed orbit $\Gamma$ and in the absence of any time-dependent applied field, relations (7a) and (7b) reduce to

$$\oint_\Gamma \left[\chi g(\mathbf{m}\cdot\mathbf{p})[\mathbf{m}\times\mathbf{p}]\cdot\frac{d\mathbf{m}}{d\tau} - \alpha\left(\frac{d\mathbf{m}}{d\tau}\right)^2\right] = 0 \qquad (8a)$$

$$\oint_\Gamma \left[\chi g(\mathbf{m}\cdot\mathbf{p})[\mathbf{m}\times\mathbf{p}]\cdot\frac{d\mathbf{m}}{d\tau} + \alpha[\mathbf{m}\times(\mathbf{h}_{\text{eff}} + \mathbf{h}_{\text{ST}})]\cdot\frac{d\mathbf{m}}{d\tau}\right] = 0 \qquad (8b)$$

The first expression states that, on a closed orbit, the work of the spin-torque equivalent field is exactly cancelled by a definite-positive damping integral and, thus, allows for the existence of precessional states under the sole action of a steady current (i.e., in the absence of any time-dependent field or current). The second establishes a relation between the work of the spin-torque equivalent field and the work of the damping field without lending itself to a clear physical interpretation as noted earlier [19], especially if damping is truncated to the sole effective field. Because, however, proceeding from the very same premises, relations (8a-b) cannot bear different meanings.

Incidentally, the sum of the exchange, magnetostatic and anisotropy energy may rise during part of the orbiting motion. If it does so, part of the energy is given back to the system at a later stage along the orbit in such a way that when integrated over one cycle, Eq. (8a) or (8b) remains satisfied. Similar situations are common, for example, in the changes in domain wall structures during wall motion under the action of a pulsed field. The energy of the domain wall increases as the wall distorts in response to the changed field. This increase is usually obscured because the micromagnetic energy as a whole decreases with time, primarily due to the decrease of the applied field (or Zeeman) energy during the wall motion. At the same time, wall distortion usually leads to an increase of the other energy components. At the end of the pulse, the wall reverses to a more stable configuration and releases the stored energy, which causes additional wall motion, called overshoot. When an applied field causes domain wall distortion and propagation, that field provides the initial torque; when a spin-polarized current causes distortion, oscillation and/or propagation, the magnetization distribution is set into motion thanks to the initial torque provided by the current.

Now, precessional states become stable in nano-pillars for values of $\chi/\alpha$ close to unity. Therefore, relaxing the spin-torque contribution to damping is equivalent to the neglect of a term of order $\sim \alpha^2$ in the damping process. Numerical simulations confirm that, when dealing with precessional states in either of the geometries considered here, the Gilbert form or the Landau-Lifshitz approach with the conservative field as a sole source of damping lead to virtually undistinguishable results.

(iii) The damping process may be affected by "spin-pumping" [28, 29], i.e. a transfer of angular momentum via the inelastic scattering of electrons with energies close to the Fermi energy flowing from the ferromagnet into the normal metal spacer or lead. Practically, this could be taken into account via a specific surface damping mechanism within numerical micromagnetics. To the knowledge of the authors, surface damping effects have not yet been included into micromagnetic simulations.

(iv) Temperature is most commonly introduced via a stochastic thermal field leading to what is commonly termed Langevin magnetization dynamics, following the pioneering work of Brown [30], where this method is introduced for a single magnetic moment. The justification of this approach within numerical micromagnetics, where several interactions between elementary moments (discretization cells) do exist, is discussed in detail in [31]. We note also that in systems submitted to spin-transfer torques, an additional source of noise exists, namely spin current fluctuations.

(v) The spin-torque equivalent field may be supplemented with an additional term that appears from first principle calculations [22] and is linked to the imaginary part of the mixing conductance in circuit theory (see [32] and Ref. therein). This additional field may, for obvious geometrical reasons, only be parallel to **p**. The total spin-torque equivalent field then reads $\mathbf{h}_{\text{ST}} = \chi \left[ g(\mathbf{m} \cdot \mathbf{p})[\mathbf{m} \times \mathbf{p}] + \beta \mathbf{p} \right]$, $\beta \ll 1$. The so-called "non-adiabatic" spin-transfer term introduced in order to explain wall motion in nano-wires at low current densities [33, 34] is phenomenologically similar. The inclusion of such an additional term will not be considered below (simulations performed by one of the authors in the past have shown that for the problems of interest here, addition of a $\mathbf{h}_{\text{ST}}^{\text{add}} = \chi \beta \mathbf{p}$ does not significantly impact simulation results).

## III. Validity range of the macrospin approximation

### III.1. Spin-torque driven dynamics in the macrospin approximation

Consider an elliptical soft magnetic element deprived of any growth-induced anisotropy. Shape anisotropy tells us that the long axis of the ellipse is the easy magnetization axis and that moving the magnetization out of the plane will prove more costly than moving the magnetization away from the easy axis in the plane. The energy functional of such an element may be reduced to:

$$\varepsilon = K\left(1 - m_x^2\right) + \frac{1}{2}\mu_0 M_S^2 m_z^2 - \mu_0 \mathbf{M} \cdot \mathbf{H}^{\text{app}} \qquad (9),$$

where $\hat{x}$ coincides with the easy axis and $\hat{z}$ is normal to the plane of the element. The first term in (9) is the uniaxial anisotropy ($K > 0$; $K \ll \mu_0 M_S^2$); the second term describes demagnetizing effects in the thin film limit, the third is the Zeeman energy. If the external field is applied along the easy axis with a sole component $H_x^{app}$, then the effective field for this element treated in the single spin limit reads:

$$\mathbf{H}_{eff} = -\frac{1}{\mu_0 M_S}\frac{\delta \varepsilon}{\delta \mathbf{m}} = \left[H_x^{app} + H_K m_x, 0, -M_S m_z\right] \quad (10),$$

where $H_K = 2K/\mu_0 M_S$ is the anisotropy field. Let us further assume the spin torque to be symmetrical ($\chi_a = 0$, i.e. $g(\mathbf{m}\cdot\mathbf{p})=1$) and the electron polarization to be aligned with the ellipse long axis ($\mathbf{p}=\hat{x}$). Solving (1) or (2) augmented with (3), that is with spin-torque added, soon leads to a phase diagram such as shown in Fig. 1 ($T=0$).

As long as $H_x^{app}$ remains smaller than the anisotropy field, there exists a transition between the parallel "P" state (both the "pinned" and "soft" layers are magnetized along the $+\hat{x}$ direction) and the anti-parallel "AP" state ("pinned" magnetized along $+\hat{x}$, "soft" layer along $-\hat{x}$) for a positive current (the convention here is that the current is positive if flowing from the "pinned" to the "soft" or "free" layer; stated otherwise, the current is negative if electrons flow from the "pinned" to the "free" layer). Conversely, the transition occurs between the "AP" and "P" states for negative currents. However, the transition is not direct, except for pathological points $\chi/\alpha = -1/2$; $H_x^{app} = +H_K$ and $\chi/\alpha = 1/2$; $H_x^{app} = -H_K$; sustained precession states are expected from the simple theory with "clamshell"-type orbits for current densities $\chi_1 < \chi < \chi_2$, and out-of-plane orbits for $\chi > \chi_2$.

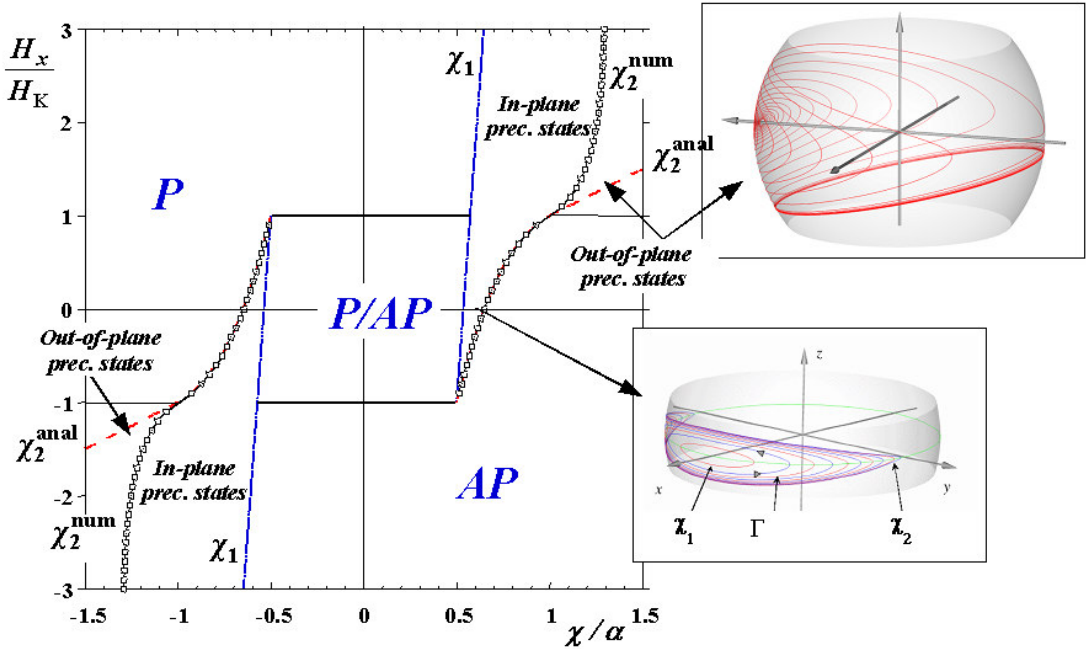

Fig. 1. Simulated phase diagram in the single spin limit at zero temperature as a function of reduced coordinates $\chi/\alpha$ and $H_x/H_K$. The current is negative when electrons first cross the "Pinned" and, then, the "Free" layer. The phase diagram separates regions where either the "P" or the "AP" state are stable from regions where, either bi-stability exists "P/AP", or sustained precession under the sole action of a steady current. The single spin limit establishes a clear difference between "in-plane" or "clamshell" orbits and "out-of-plane" orbits to be found at larger current densities for fields in excess of the anisotropy field (Upper-right and lower-left regions of the phase diagram delimited by the horizontal segment $H_x/H_K=1$ or $-1$ and the computed phase boundary $\chi_2^{num}$). Blue and red lines are computed analytically (see [23]) whereas open symbols are simulation results.

Our main point here is to show that an actually terribly over-simplified macrospin approach does lead to a phase diagram for an applied field along the easy anisotropy axis that has much in common with the experimental phase diagram to be found in, e.g., [35] (except for the field offset observed in the experimental phase diagram due to the magnetostatic coupling between the "soft" and "hard" layers). Three additional remarks here: (i) rounding-up of the phase diagram boundaries is expected from finite temperature effects as well as regions of states overlap [36], (ii) precessional states have been observed not only for $|H_x^{\mathrm{app}}| > H_K$, but also in the region $|H_x^{app}| < H_K$ [37, 38], (iii) the phase diagram in Fig.1 proves only mildly robust in the presence of a transverse in-plane field component even at zero temperature [39].

As recalled above, the single-spin approximation is able to capture some fundamental aspects of spin-torque induced (STI) magnetization dynamics. In some instances even analytical calculations may be pursued far enough so as to define thresholds for the onset of precessional states and switching. However, recent experiments [40] have demonstrated that switching in these systems proves complex, a fact actually predicted by all full micromagnetic simulations previously performed. It ought to be mentioned, however, that critical currents are particularly high in [40], resulting in an exalted influence of the Oersted field, i.e., the field created by the current flowing across the pillar. In order to understand why the macrospin approximation may fail, although the size of many systems where the STI magnetization dynamics has been observed is very small, we first recall several important issues concerning the determination of critical single-domain sizes.

### III.2. Critical sizes for the validity of the macrospin approximation

First we remind the reader that the concept of a 'strictly' single-domain magnetic particle (body) does indeed exist. We mean here a particle that remains homogeneously magnetized independently of external conditions, like the value and the direction of the applied field (of course, the temperature is still assumed to be well below $T_c$). The qualitative estimation of the critical size for such a 'strictly' single-domain behaviour relies on the comparison between (*i*) closed-flux magnetization configurations governed by a sole exchange energy $E_{\mathrm{exch}}$ and (*ii*) the collinear state characterized by $E_{\mathrm{exch}} = 0$ and $E_{\mathrm{dem}} > 0$, where $E_{\mathrm{dem}}$ is the magnetostatic energy (see Chap 3.3 in [4] and refs. therein). The exchange energy density of the closed $\mathbf{M}(\mathbf{r})$-configuration obviously increases with decreasing particle size (magnetization gradients are getting larger), leading to the result that only a collinear magnetization state is energetically stable below the critical size $l_{\mathrm{cr}}$ which scales with the exchange length $l_{\mathrm{exch}} = \sqrt{2A/\mu_0 M_S^2}$. We note in passing that the anisotropy energy is not included into these considerations; in general, for a single crystal ferromagnet, this anisotropy is expected to stabilize the homogeneous magnetization state, thus increasing $l_{\mathrm{cr}}$. An exact value of $l_{\mathrm{cr}}$ can only be determined by rigorous numerical simulations. It depends on many physical factors, and amongst them, primarily the particle shape. It also depends on the non-collinear state used to compute $E_{\mathrm{exch}}$, and, in most cases, amounts to $l_{\mathrm{cr}} \approx (4-8) \cdot l_{\mathrm{exch}}$ [4]. The exchange length is about $l_{\mathrm{exch}} \approx 5$ nm or less for most ferromagnets, so that an *upper* bound for the single-domain threshold is $l_{\mathrm{cr}} \approx 40$ nm. This would mean that for nearly all spin-torque experiments (leaving aside data obtained on point contacts to be discussed below) the lateral element size is far above $l_{\mathrm{cr}}$, invalidating the macrospin approximation and raising the question why this approximation can give any reasonable predictions at all. In order to answer this question in particular and to make further methodological progress in general, several factors need to be taken into account:

(i) The $l_{\mathrm{cr}}$-estimate given above is solely based on the *energy* comparison between different

configurations. Thus it *can not* be used to predict whether the transition from a single- to a multi-domain state will *really* occur during a remagnetization process, because this transition often requires overcoming an energy barrier. This means, that an approximately homogenous (single-domain) magnetization state, being for some specific external conditions only metastable, can still exist, because the transition to, e.g., some closed magnetization configuration with a smaller energy requires overcoming a prohibitively large energy barrier. Indeed, simulations have shown that for certain particle shapes almost collinear magnetization states persist during the whole remagnetization process for nanoelements with lateral sizes as large as several hundred nanometers in a *homogeneous* external field.

(ii) Most calculations leading to the estimation $l_{cr} \approx (4-8) \cdot l_{exch}$ given above were performed for particles with sizes of the same order of magnitude in all three dimensions (cubes, spheres etc.). For a thin film element with thickness much smaller than its lateral sizes, both exchange and stray field energies might have a different size dependency, which, in turn, might substantially affect $l_{cr}$.

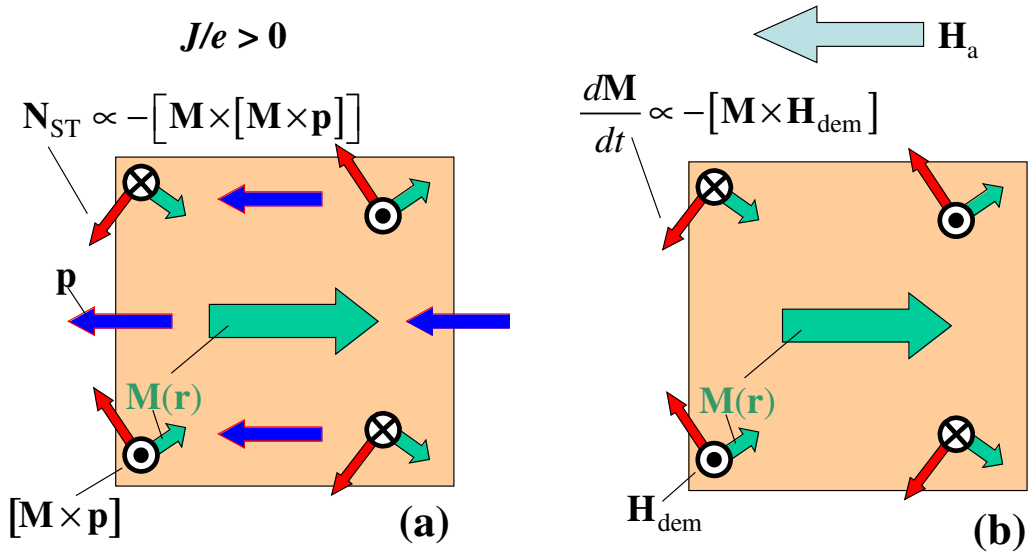

Fig. 2. (a) Torque distribution in a nearly uniformly magnetized particle ("*flower*" state) for a current promoting parallel alignment ($J$, $e$ < 0 in Eq. (4)): the torque is strongly inhomogeneous even for a slightly non-collinear magnetization configuration; (b) Torque distribution under the action of a homogeneous applied field.

(iii) Looking at the spin-torque distribution in quasi-single domain elements proves instructive. For, e.g., a square nanoelement in a "*flower*" remanent state and for a spin polarization collinear with its mean magnetization direction, the spin torque (4) has *opposite* directions near adjacent corners of the square as displayed in Fig. 2a. In other words, the initial spin torque distribution proves highly non-homogeneous [41], impairing easy magnetization reversal. Such a situation is however not unique to spin-torque action on a non uniform magnetization distribution: in a thin film element, reversal under the action of a homogeneous field antiparallel to the mean magnetization direction would result in an equivalent initial torque distribution since the precession of the magnetization around the applied field gives rise to a local demagnetizing field $\mathbf{H}_{dem}$, with, as a result, magnetization precession around the latter. The cross-product $[\mathbf{M} \times \mathbf{p}]$ actually plays the role of $\mathbf{H}_{dem}$.

The line of arguments presented above means that *each specific experimental situation*

requires a separate analysis using full-scale micromagnetic simulations to find out whether the macrospin approximation is valid for its description. An example of such an analysis can be found in [42], where spin-torque induced (STI) precession in a square element with thickness $h = 2.5$ nm and typical magnetic parameters[2] ($M_S \approx 1000$ G, A = $2 \cdot 10^{-6}$ erg/cm) was studied. It has been shown than the magnetization configuration during STI precession significantly deviates from single-domain behaviour already for lateral sizes as small as $L = 30 - 40$ nm. This value is significantly smaller than the lateral sizes of nanoelements used in all experiments reported so far in nanopillars.

Concluding this discussion, we point out that there exists a class of experimental systems where the macrospin approximation is *invalid* for *any* sizes of the current flooded area, namely the so called point-contact setup. In such experiments [43, 44, 45, 46], a current-carrying wire with a very small diameter (20 - 80 nm) is attached to a system of *extended* magnetic layers, usually with lateral extension ~ 10 μm. As mentioned above, the size of the area flooded by a current through the 'free' magnetic layer can be as small as 25 nm [43], making it very tempting to declare that the macrospin approximation is applicable to analyze these experiments even quantitatively.

However, in contrast to the multilayer nanopillars, where the whole area of magnetic layers is flooded by the electric current, in the nanocontact setup only the magnetization within the area under the contact 'feels' a spin-polarized current and can be directly excited by it. On the other hand, there exists a strong exchange interaction between this area and the outer film region (which results, in particular, in the necessity for higher current densities in order to excite steady-state magnetization precession when compared to the nanopillar geometry). This exchange interaction is qualitatively important: first, it should be included in order to accurately determine the equilibrium magnetization configuration of the system; it may be quite complicated due to a large Oersted field arising for the high-density currents through the contact. Second, exchange is responsible for the formation of spin waves that carry energy away from the point contact area, so that the inclusion of this interaction is crucial even for a qualitative understanding of the magnetization dynamics [20, 47, 48]. Unfortunately, within the macrospin formalism there is no adequate method to incorporate exchange. Hence, the macrospin model is, strictly speaking, invalid in this situation for any size of the point contact.

### III.3. Misleading artefacts of the macrospin approximation

As briefly recalled above, the macrospin dynamics is very rich, so that the macrospin model can successfully reproduce many features of the actual magnetization dynamics of real systems. The reverse of this coin is the danger that due to this richness the macrospin approximation can accidentally reproduce some real dynamics features suggesting a completely misleading explanation of them.

An excellent example of such an artefact can be found in the famous paper of Kiselev et al. [49]. Trying to explain many interesting and highly non-trivial features of the experimentally observed STI dynamics in Co/Cu/Co nanopillars, Kiselev et al. performed macrospin simulations. Using several adjustable parameters (like the saturation magnetization and an homogeneous uniaxial anisotropy), they succeeded in reproducing not only the value of the magnetization oscillation frequency and the correct slope of its current dependence, but also the frequency jump with increasing current ! Within the framework of macrospin simulations this jump was elegantly interpreted as the transition between the small amplitude elliptical trajectory and the large amplitude 'clam-shell' trajectory (see Fig. 2 in [49]), so that a nearly perfect quantitative agreement between experiment and macrospin simulations was achieved.

---

[2] Most material parameters are expressed in cgs units because these units are still used by most experimentalists. Use the following transformations if necessary: 1000 G is equivalent to $10^6$ A/m; in other words, 1 kA/m (SI) is equivalent to 1 G (cgs). 1 Oe is equivalent to $(1000/4\pi)$ A/m. A field equal to 1 mT means a 10 Oe field. Exchange constant of $10^{-6}$ erg/cm is equivalent to $10^{-11}$ J/m. Lastly, the energy density (anisotropy constant) of 1 erg/cm$^3$ is equivalent to 0.1 J/m$^3$.

Unfortunately, full-scale micromagnetic simulations performed later [50] have revealed that this jump is an artifact of the macrospin approach. For the elliptical nanoparticle with the geometry given in [49] and magnetic parameters typical for Co, transition from regular magnetization oscillations (where the nanoelement magnetization remains at least approximately collinear) to a "chaotic" magnetization dynamics (which is a signature of a highly irregular magnetization configuration during oscillations) occurs before the transition from 'small-angle' to 'large-angle' oscillations, so that the corresponding frequency jump can not be explained correctly by the macrospin model. We shall return to the analysis of this very challenging paper later on.

## IV. Steady-state precession in the nanopillar geometry

In this section we focus our attention on micromagnetic simulations of a steady-state precession in nanopillars, i.e., persisting magnetization oscillations when a dc-current flows through a multilayer stack of thin magnetic nanoelements separated by non-magnetic spacer(s). We actually concentrate mainly on two such experiments [49, 51]) for several reasons. First, both papers contain highly non-trivial results (and for this reason are probably the most frequently cited on this topic). Second, both studies include a reasonable [49] or very detailed [51] sample characterization, so that attempting a quantitative comparison with simulation data makes sense. Third, systematic numerical simulations have been performed in order to understand dynamics observed in [49] (see, e.g., [52, 50, 53, 54]) and very recently [55, 39], also to analyze results from [51].

Before we present detailed analysis of these two experiments, let us point out that, apart from the use of different magnetic materials, the experimental setups used in [49] and [51] differ in two fundamental aspects. In [49], both the applied field and the conduction electron polarization are nominally collinear and pointing approximately along the long axis of the elliptical soft element. We shall refer here to a "longitudinal" geometry. In [51], the angle between the field and the electron polarization is close to 90° and neither the field nor the electron polarization direction coincide with the major ellipse axis. We shall call this setup a "skewed" geometry. As demonstrated below, the external field direction as well as the electron polarization direction are qualitatively important for the STI-dynamics, so that we analyze these two experimental situations separately.

### IV.1. Longitudinal geometry : STI-dynamics features which can be explained by simulations

We start with the analysis of the famous paper of the Cornell group [49], which was historically the first one allowing for a quantitative analysis of experimental results on STI-dynamics. In this work, the authors used a Co/Cu/Co nanopillar structure with an extended bottom thick Co layer (40 nm) and elliptical thin Co (3nm) nanoelement with nominal lateral size 130 nm x 70 nm on top of it. Classical magnetostatic coupling between the "hard" and "soft" layers was thus presumably moderate if not absent. Furthermore, since the non-magnetic Cu spacer was 10 nm thick, any indirect exchange interaction between the two Co layers could be excluded. When the electron flow from the dc-current was directed from the thin to the thick Co layer, Kiselev at al could detect microwave-frequency oscillations of the sample resistance in a large interval of currents and external magnetic fields. As mentioned above, in the Kiselev experimental geometry the magnetization direction of the 'fixed' layer and the applied field are parallel and nearly aligned with the long axis of the elliptical 'soft' ('free') element. The observed dynamics had the following main features:

1). Small-amplitude signal at low currents with a (high) frequency virtually independent from the current

2). Huge frequency drop (from ≈ 17 GHz to ≈ 6.5 GHz) when the current was increased, accompanied by a dramatic growth of the signal amplitude ('large-amplitude' signal) at first, followed by a vanishing-out of the rf power for the largest currents (≈ 6 mA).

3). Continuous frequency decrease with increasing current in the large-amplitude regime.

4). The large-amplitude signal showed several equidistant spectral bands, where upper bands were obviously higher harmonics of the basic frequency $f_0$ and a significant power contribution in the regime of very low (compared to $f_0$ !) frequencies (0 – 1 GHz).

5). Broad spectral lines (~ 1 GHz), especially in the large-amplitude regime.

A quantitative analysis of the system magnetization dynamics using micromagnetic simulations proved far from straightforward, due first to the large sample-to-sample variations of experimental data and, second, to rather controversial specifications of the Co magnetic parameters given in [49]. However, many important features listed above could be successfully reproduced at least qualitatively [50], namely:

(i) *Small-amplitude signal with the current-independent frequency*. According to simulations, the 'small-amplitude' signal with a nearly constant (current-independent) frequency corresponds to the small-amplitude (linear) magnetization oscillations excited for currents only slightly different from the threshold current, $j_{cr}$, for the steady-sate precession onset at zero temperature. When the current is not much higher than $j_{cr}$ and increases, then the oscillation amplitude increases also, but remains still small enough to ensure a quasi-independence of the oscillation frequency from oscillation amplitude, as expected from standard classical mechanics. An important point here is that as long as thermal fluctuations were neglected in simulations, the oscillation amplitude was found to grow very rapidly with the current $j > j_{cr}$ (so called 'stiff' oscillation generation), so that the current interval where the amplitude remained small enough to keep the frequency almost constant, was negligibly small. Only inclusion of thermal fluctuations allowed for a gradual (relatively slow) increase of the oscillation amplitude, so that the 'small-amplitude' regime could be satisfactory explained [50].

(ii) *Downwards frequency drop with increasing current*. This jump was interpreted by Kiselev et al. within the framework of macrospin simulations as the transition between the small-angle elliptical precession and large-angle 'clam-shell' orbit motion of the macrospin. Numerical simulations [50] revealed that the coherence of the magnetization configuration is completely lost *before* the transition to the 'clam-shell' orbit, so that the macrospin picture is invalid for such large currents. These simulations have shown that the transition to fully incoherent magnetization oscillations (different parts of a nanoelement oscillate with different frequencies and amplitudes, leading to the so called quasichaotic regime) is accompanied by the abrupt decrease of the oscillation frequency. This abrupt decrease could be interpreted as a frequency jump in the experiment. However, the amplitude of the jump found in simulations ($\Delta f_{exp} \approx 3$ GHz) proved by far not as large as in the experiment (($\Delta f_{exp} \approx 10$ GHz).

(iii) *Frequency decrease with increasing current in the 'large-amplitude' regime*. Increasing the current strength normally leads to an increase of the oscillation amplitude, simply because more energy is 'pumped' into the system. In principle, the reasons why the frequency may strongly depend on the oscillation amplitude in the case of non-linear oscillations – and oscillations in the large-amplitude regime are obviously non-linear - is well known in mechanical systems. A detailed explanation of this phenomenon for magnetization dynamics in the macrospin approximation can be found, e.g., in [47]. Roughly speaking, the oscillation frequency $f$ is proportional to the product $f \propto \sqrt{H(H + N_d M_{eq})}$, where $M_{eq}$ is the magnetization projection on the *equilibrium* direction of the magnetization, around which precession takes place, and $N_d$ is the demagnetizing factor. For oscillations around the in-plane orientation of the magnetization (which was the case in experiments under discussion, due to the in-plane anisotropy of a thin ellipsoidal plate and the in-plane external field), this factor is positive, and the average projection of the magnetization on its equilibrium (in-plane) direction obviously decreases with increasing amplitude. Hence the second factor under the square root $\sqrt{H(H + N_d M_{eq})}$ decreases, pushing the frequency down. In the quasichaotic regime, when the macrospin approximation cannot be applied any longer, one can still follow the same line of arguments, saying that qualitatively the relation $f \propto \sqrt{H(H + N_d M_{eq})}$ remains valid, and the average magnetization drops with increasing current simply because in the quasichaotic regime the magnetization becomes more inhomogeneous when the current strength grows, thus leading to the same effect ($f$ decreases with $I$). However, one should keep in mind, that for oscillations with the characteristic wavelengths as small as in the experiments under consideration (~ 10 – 20 nm) the contribution of the exchange interaction to the dynamical system behaviour is very significant, so that a theory of quasichaotic dynamics which includes also the exchange stiffness of the system, is actually required for a thorough understanding of its behaviour.

(iv) *Several equidistant spectral bands*. The presence of significant spectral peaks with frequencies at $2f_0$, $3f_0$, etc. where $f_0$ denotes the 'basic' oscillation frequency, is usually considered in classical mechanics as an evidence that the oscillations are strongly non-linear, so that the time dependence of the oscillating co-ordinate is no longer simply sinusoidal and hence its Fourier expansion contains substantial contributions at frequencies corresponding to higher harmonics. When we consider magnetization dynamics, this argument remains valid; however, an additional complication arises due to the fact that magnetization oscillations actually arise from the *precession* of the magnetization vector. This precessional character of the magnetization oscillation leads to the following picture.

When the magnetization vector **M** with constant magnitude $M_0$ oscillates around, say, the $x$-axis with frequency $f_0$, then its $y$- and $z$-projections oscillate with the same frequency: $M_y = M_0 \sin\theta_0 \cos(2\pi f_0 t)$, $M_z = M_0 \sin\theta_0 \sin(2\pi f_0 t)$, where $\theta_0$ is the precession angle, i.e., angle between **M** and the x-axis. In the simplest case, assuming circular precession, $\theta_0$ is constant (time-independent) and so is the $x$-projection $M_x = M_0 \cos\theta_0$. However, in the experimental situations under consideration, the shape anisotropy of the magnetic nanoelement gives rise to an elliptical orbit and $M_x$ is also time-dependent: $M_x = M_0 \cos\theta_0(t)$. Now, because the magnetization magnitude should be kept constant, the elliptical precession of $\mathbf{m} = \mathbf{M}/M_s$ takes place on the unit sphere and the angle $\theta_0$ reaches its minimal (and its maximal) value *twice* during one oscillation period. This means that the $x$-projection of the magnetization oscillates with the frequency $2f_0$ even when the precession angle is small, i.e. well in the linear regime.

To proceed further, we note that magnetization oscillations are detected experimentally using some kind of a MR-effect (here the GMR), which is proportional to the scalar product of magnetizations of the 'fixed' and 'free' layers: $\Delta R(t) \propto \mathbf{m}_{\text{free}}(t) \cdot \mathbf{m}_{\text{fixed}}$. For the in-plane external field (the case studied in [49]), the fixed layer moment lies in the nanoelement plane, which we denote as the *xy*-plane. Then the MR time dependence is given by $\Delta R(t) \propto m_{\text{free}}^{(x)}(t) \cdot m_{\text{fixed}}^{(x)} + m_{\text{free}}^{(y)}(t) \cdot m_{\text{fixed}}^{(y)}$, thus containing both the basic frequency $f_0$ coming from $m_y$-oscillations, and the next harmonic $2f_0$ due to $m_x$-oscillations. This logic is valid for arbitrary small oscillation amplitude, so that in the geometry with the in-plane orientation of the external field and the fixed layer magnetization the second harmonics should be present in the linear regime also. An important exception is the case when equilibrium orientations of the 'free' and fixed layers coincide, which may be the case when either the external field is very strong or the experiment geometry has a special symmetry – like an elliptical (in-plane) multilayer stack with the external field oriented exactly along the major ellipse axis. In this case one should *not* detect the signal with the *basic* frequency $f_0$. To understand why this is so, it is enough to choose the *x*-axis along the equilibrium orientations of the magnetizations of both layers. Then $m_{\text{fixed}}^{(y)} = 0$ and we are left with the signal due to oscillations of $m_{\text{free}}^{(x)}(t)$ *only*, which have the frequency $2f_0$. The basic frequency is then not observed at all !

(v) *Large spectral linewidth in the large-amplitude regime*. Such broad spectral lines, observed for magnetization oscillations of so small elements, are normally explained as a thermal fluctuation effect. However, micromagnetic simulations have shown, that even without including these fluctuations into the equation of motion, we still observe very broad spectral lines due to the quasichaotic nature of magnetization oscillations in the 'large-amplitude' regime. On the other hand, the explanation of still relatively large linewidths in the regular regime (for current only slightly above the critical current for oscillation onset) definitely requires the introduction of thermal fluctuations [50].

(vi) *Absence of the out-of-plane regime due to the loss of coherence*. The so called 'out-of-plane' oscillation regime when the magnetization of a nanoelement is oscillating around an axis pointing out of the element plane, is an inherent feature of the steady state precession in the macrospin approximation (see above). The signature of this regime is an increase of the oscillation frequency with increasing current. Spectral bands with the corresponding *f(I)*-dependence have been observed neither in [49], nor in most other experiments carried out in the nanopillar geometry when the external field was oriented in the film plane. Simulations have explained why these out-of-plane oscillations are almost never observed: macrospin theory shows that 'out-of-plane' precession should occur for large currents and values of the external field large enough for a *single* energy minimum to exist. However, for such large currents the coherence of the magnetization configuration during the steady-state precession is largely lost, so that conditions for the existence of this regime are not fulfilled anymore. Modelling the situation studied in [49] has shown that for the currents such that out-of-plane precession would be expected from the macrospin model, relatively small fluctuations of the random crystal grain anisotropy of Co and thermal fluctuations are strong enough to significantly disturb the collinearity of the magnetization configuration. Thus the oscillation amplitude of the *average* magnetization is strongly reduced, so that the out-of-plane regime could hardly be detected experimentally [50].

Simulations based on a similar geometry, but with magnetic parameters typical of Permalloy (saturation magnetization similar to those used in [50] $M_S$ = 800 Oe, but with a much smaller exchange constant A = $1.0 \cdot 10^{-6}$ erg/cm) and a slightly reduced element size [53, 39], have revealed another possibility to explain the features of the 'small angle' regime found in [49]. Montigny et al. argued that the almost constant frequency regime with a low oscillation power should correspond to the first eigenmode of the elliptical element. For the exchange constant used in [53], it was shown that this mode is localized near the edges of the elliptical nanoelement. Upon current increase, the red-shift mode involving a quasi-uniform precession within the whole element was seen to develop with a fast initial rise in rf power. It was further found [39] that for Permalloy-like parameters, "clamshell" type orbits do indeed exist in the micromagnetic regime although rather irregular, resulting in a broad linewidth. Unlike the single spin picture, the largest opening of "clamshell" orbits was not larger than ≈ 180°, in agreement with [50].

The frequency drop at the onset of "clamshell" orbits is of the order of 2 GHz for the material parameter considered, which is similar to the value obtained in [50], but much smaller than observed in [49]. Montigny's results seem more in tune with recent experiments of Mistral et al. [56], in spite of the line width in Mistral's experiment proving significantly narrower than ever uncovered in simulations (note also that the red-shifting region in Mistral's experiments proves extremely narrow).

The existence of spectral bands is quite generic as soon as the angle between the magnetization at rest within the hard and soft layers departs from zero. Lastly, in the same elements, it was found that out-of-plane orbits also exist, with as a consequence the existence of a blue-shift regime, even in full micromagnetic simulations. However, the out-of-plane trajectories prove notoriously unstable so that the system may spend a few nanoseconds precessing out of the plane, then decay in some kind of chaotic trajectory of the mean magnetization, before eventually turning back to an out-of-plane orbit. Telegraphic noise type hopping between out-of-the plane orbits above and below the plane of the sample is not excluded.

Further discussion of simulation results is deferred until Section IV,4. At this stage, however, it appears that the most meaningful difference between simulations outputs in [50] and [53, 39] occurs in the low current regime. At higher currents, both authors conclude to the existence of highly perturbed and limited extent orbits before a "weak" out-of-plane precession regime takes over.

### IV.2. Skewed geometry (isolated free element): features of the STI-dynamics which can be reproduced and explained by simulations [55]

There are two main differences between the multilayered nanopillars used in [49] and [51]. First, the stack in [51] contains Py layers instead of Co as magnetic components and second, the lower Py layer was sputtered onto IrMn, an antiferromagnet (AFM). It is well known that when a FM layer is deposited onto a high-quality layer of some specific AFMs (including IrMn), the exchange interaction between the FM and the upper atomic layer of the AFM, a phenomenon known as 'exchange bias', stabilizes the FM magnetization in the direction parallel to the orientation of the upper AFM magnetic moments. Krivorotov et al. have used this phenomenon to 'pin' the magnetization of the lower Py layer in a direction significantly tilted away from the long axis of the 'soft' elliptical element. Because the equilibrium magnetization directions of the 'free' and 'fixed' layers are clearly *not* collinear (look at geometry in Fig. 5), the initial torque exerted by the spin-polarized current is anticipated to be almost homogeneous. For this very reason, the whole structure is expected to display a markedly more coherent behaviour [41].

An additional advantage of the experiment reported in [51] was the careful characterization of the Py layers. Not only the saturation magnetization $M_S$ (which turned out to be quite low - $M_S \approx 640$ G), but also the magnetization damping parameter $\alpha$ (= 0.025) entering the LLG equation of motion for the system magnetization were measured independently - see [51, 55] for details. Another simplification occurs due to a very low value of the Py magnetocrystalline anisotropy ($K \approx 5 \cdot 10^3$ erg/cm$^3$), so that this anisotropy and thus - the grain structure of Py can be safely neglected. Finally, the experiments under discussion were carried out at low temperatures (nominally at liquid He, but due to the Joule heating the actual temperature was estimated to be $T \sim 10 - 50$ K), so that thermal fluctuations should be less important than for the data from [49], hopefully leading to an easier analysis.

As for the previous experiment, we first list the most important features of the experimental results [51]:

1) Strong decrease of the oscillation frequency with increasing current

2) Existence of several frequency jumps with increasing current

3) Evidence for extremely narrow line widths ($\Delta f \sim 10 - 100$ MHz), which vary non-monotonically with current.

Now we turn our attention to the analysis of these features based on the insight offered by numerical simulations.

(i) *Decrease of the oscillation frequency with increasing current strength*: The frequency drop observed in [51] proves quite different from the 'red' frequency shift discussed previously: the frequency decreases very rapidly from the very beginning of the oscillation onset, exhibiting several jumps. We remind that the frequency of 'small-amplitude' oscillations observed in [49] was nearly current-independent up to the transition to the 'large-amplitude' regime (i.e., in the region 1.7 - 2.4 mA), after which the frequency started to decrease continuously with increasing current (without further jumps). Numerical simulations have shown, as expected, that the rapid frequency decrease immediately after the oscillation onset is a non-linear effect due to the fast growth of the oscillation amplitude with increasing current. This increase of the oscillation amplitude could also be detected as a fast increase of the measured microwave oscillation power emitted by the device [55]. Due to the careful experimental characterization of the samples numerical simulations could reproduce this fast initial frequency decrease not only qualitatively, but also *quantitatively* without any adjustable parameters (See Fig. 3).

(ii) *Origin of the frequency jumps*. The nature of the frequency jumps in [51] turned out to be also very different from that of the jump observed in [49]. The jump in [49] signalled the transition from the regular (with a relatively homogenous magnetization configuration) to the quasichaotic oscillation regime owing to Berkov's analysis. This reason could be safely excluded for the system studied in [51], because both *before* and *after* the jumps, the oscillation linewidth was extremely narrow (becoming somewhat larger in the immediate vicinity of the jumps). We could show that the frequency jumps in this case were due to the transitions between different types of regular (non-chaotic) oscillation modes, namely - between modes with different *spatial* localizations.

Simulations have revealed the following picture. At the beginning of the magnetization oscillations spatial power is relatively homogeneously distributed across the nanoelement. When the current is so large that the oscillation amplitude of these homogeneous oscillations is nearly maximal, the transition to a mode localized in the central region of the elliptical nanoelement occurs. This transition results in the first frequency jump, which was observed for many samples studied by the Cornell group. For some samples - including that studied in detail in [55] a 2$^{nd}$ frequency jump could also be observed. Again, simulations have shown that this jump is due to a more subtle transition - namely, from the mode localized only in the direction along the major ellipse axis to the mode localized in all directions (see corresponding detailed explanation with spatial maps of the oscillation power for various localization types in [55]). This 2[nd] transition was observed experimentally for only a few samples; taking into account that simulations were performed on the structure with perfect borders, we can assume that samples where this transition was found, have an especially neatly-shaped edge surface.

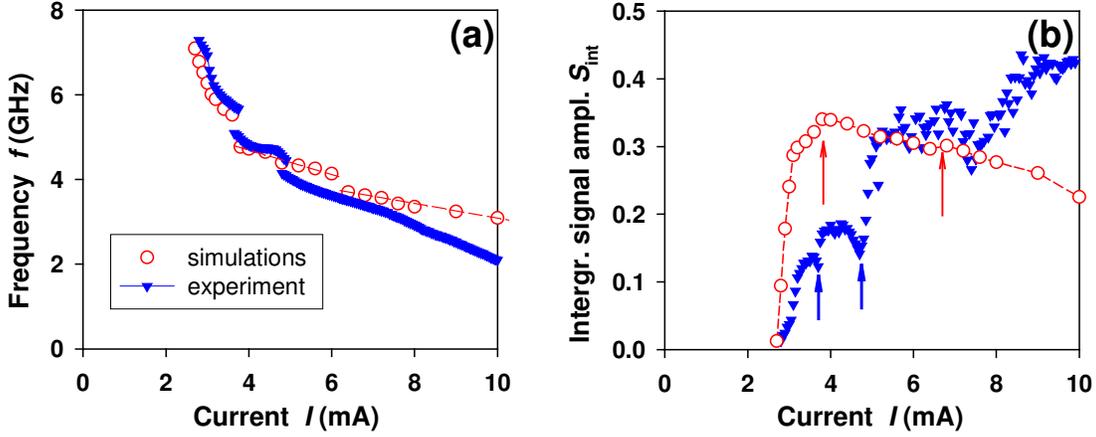

Fig. 3 (a) Comparison of experimentally measured (triangles) and simulated (circles) dependencies of the oscillation frequency of on current strength. (b) The same for the integrated spectral density $S_{int}$.

(iii) *Very small spectral linewidth*, A linewidth as small as 10 MHz for ranges of current values is a reliable evidence that magnetization oscillations observed in [51] remain regular up to the currents about at least ~ 8 mA. This is also in sharp contrast with results in [49], where the transition to the quasichaotic behaviour occurred already for $I \approx 2$ mA for the nanoelement with approximately the same lateral sizes (130 x 70 nm$^2$). This is especially astonishing, taking into account that Co used in [49] as a material for FM layers, has a much higher exchange stiffness constant ($A_{Co} \approx 3 \cdot 10^{-6}$ erg/cm) than Py ($A_{Py} \approx 1.3 \cdot 10^{-6}$ erg/cm), and a higher exchange stiffness should obviously stabilize a more uniform magnetization structure, thus preventing the system from 'sliding to chaos' (we remind the reader that according to the measurements reported in [49] and [51], we have used for Co ($M_S \approx 800$ G) and Py ($M_S \approx 650$ G) values of the saturation magnetizations which are quite close). A somewhat higher thickness of the FM layer used in [51] ($h_{Py} \approx 4$ nm) compared to [49] ($h_{Co} \approx 2.5$ nm) can hardly overcompensate such a difference in the exchange constants.

We assume [55] that the preservation of the regular oscillation regime up to such high currents as observed in [51] is due to the large angle between the polarization direction of electrons responsible for the spin torque effect and the equilibrium magnetization of the 'free' layer. Indeed, if the electron polarization vector **p** is tilted relative to the average magnetization **M** of the free layer strong enough, the spin torque **N**$_{st}$ has roughly the same direction for the magnetization across the whole elliptical nanoelement, thus supporting - and not destroying, as shown in Fig. 2a, the homogeneous magnetization state. This is most probably the reason why the regular oscillation regime 'survives' up to pretty high currents in nanopillars with the hard layer pinned at some fair angle from the easy axis.

### IV.3. Skewed geometry: simulations including the magnetodipolar interlayer interaction

Up to this point, the free layer has been treated as isolated. However, in a stack such as described in [51], the etching process is anticipated to embed both the free and the pinned layer. It ensues that the latter exerts a stray field on the former. That field is by no means small. Fig. 4 shows the values of the *x* and *y* components of the stray or biasing field averaged over the volume of the free layer as well as the extremum value of the *z* component as a function of a hypothetical cone angle of the pillar (such effects are rather common in nano-fabrication). Clearly, the average in-plane stray field components prove little sensitive to the cone angle. In contradistinction, the maximum (and minimum) value of the out-of-plane stray-field component decreases sharply with increasing cone angle although in all cases remaining particularly large. Being highly inhomogeneous, the stray field potentially affects mode localization. Moreover, this field is in no way much smaller than the applied field or the demagnetizing field from the free layer itself.

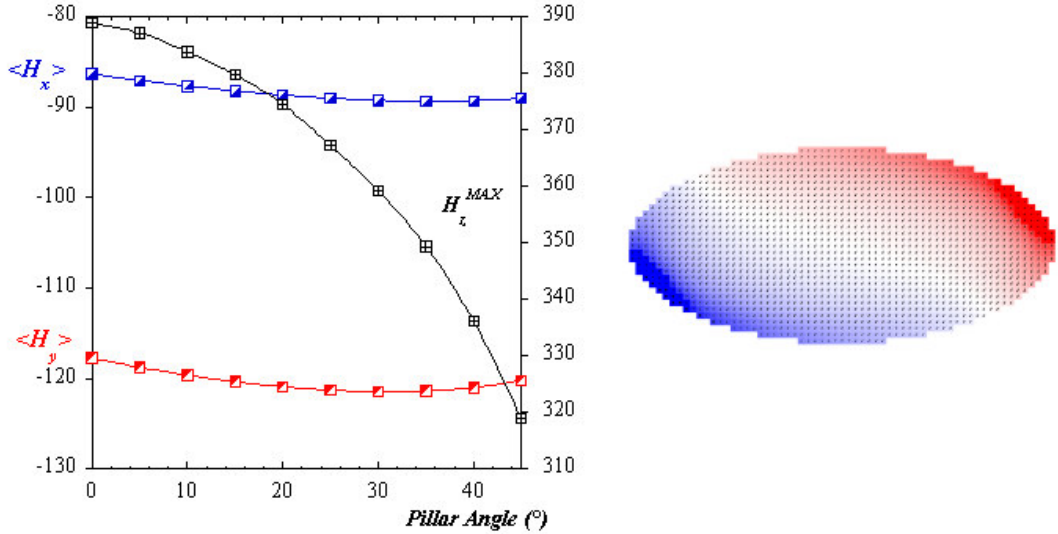

Fig. 4. Variation of the average in-plane and maximum out-of-plane components of the stray-field due to the fixed layer if pinned at 30° from the ellipse long axis. The extremum values of the out-of-plane stray-field component are to be found in the dark blue and red regions of the field map on the right

The stray-field in Fig. 4 has been corrected for stair-case effects that arise as soon as the charge distribution along the rim of the element becomes non-monotonic and piece-wise constant due to the tile decomposition of the curved rim boundary. The stray field distribution indeed becomes irregular in the immediate vicinity of a stair-case boundary, a clearly unphysical result (see [57, 58, 59, 60] for different approaches to this problem). The same remark applies to demagnetizing field computations for all simulation results in this section. Dispersion curves computed in the single spin limit for a perfectly cylindrical pillar, including the stray field ($\langle H_x^{\text{str}} \rangle = -8.63$ mT, $\langle H_y^{\text{str}} \rangle = -11.77$ mT), applied field ($H_x^{\text{app}} = +48$ mT, $H_y^{\text{app}} = -48$ mT, $H^{\text{app}} = 68$ mT) and anisotropy field representative of a 4 nm thick free layer ($M_S = 650$ kA/m) with elliptical cross-section and dimensions $2a = 130$ nm, $2b = 60$ nm (i.e., $H_K \cong 50\ mT$) is shown in Fig. 5 for two values of the damping constant: $\alpha = 0.010$ and $\alpha = 0.025$. The magnetization direction within the fixed layer is still supposed to be pinned at + 30° from the long ellipse axis.

For the small damping constant, red-shifting proves slow after the fast initial decrease because of the competition between orbit opening as a function of increasing current density and a high mobility along the trajectory due to the low damping constant. The existence of a large region with reduced *df/dJ* is reminiscent of similar features in systems with $\mathbf{p} = \hat{x}$ when a transverse in-plane field adds-up onto the usual longitudinal field [39] and is consistent with the modification of the energy landscape: the combination of the applied and biasing fields indeed acts as a field with a large longitudinal but even stronger transverse component. For the larger damping constant, there exists a critical current above which precession shifts into the out-of-plane regime, as also observed in the micromagnetic regime in the absence of Oersted field [55]. Experimental current values do not exceed 10 mA, values for which the precession frequency owing to the single spin approximation remains particularly large w.r.t. experimental values.

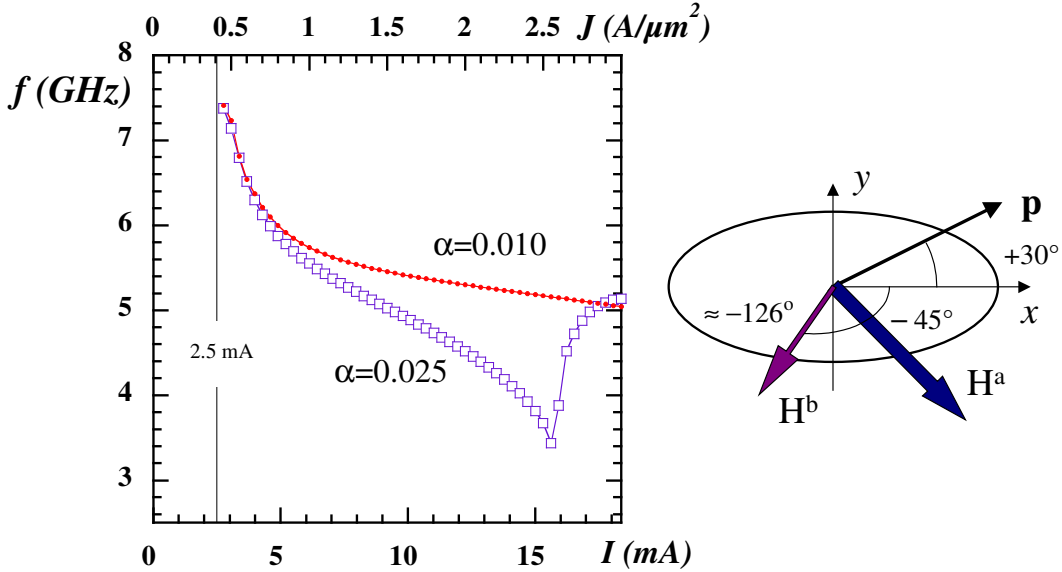

Fig. 5: Dispersion relation in the *macrospin approximation* for a "symmetrical" spin-valve Py (4 nm)/Cu (8 nm)/Py (4nm) for an asymmetry parameter $\chi_a = 0.5$. Simulation parameters: $M_S = 650$ kA/m, $H_K = 50.2$ mT, $H^{app} = 68$ mT, $H^{str} \approx 14.5$ mT The effective polarization has been adjusted so that the critical current is in both cases close to 2.7 mA, i.e. $P_{eff} = 37.5\%$ for a = 0.025 and $P_{eff} = 15\%$ for a = 0.010. The temperature is assumed equal to $T = 40$ K, independently of current amplitude.

Moving to the micromagnetic regime leads to the dispersion curves shown in Fig. 6. From these figures, several conclusions may be drawn:

(i) The thermal frequency of the order of $\approx 6.5$ GHz remains close to experimental values;

(ii) Red-shifting always occurs faster for simulations operated in the full micromagnetic regime as compared to macrospin simulations, which, as noticed previously, are unable to predict mode hopping;

(iii) Contrary to simulations performed in the absence of any biasing field, true mode hopping is only observed for the low damping constant, whereas, for the higher value, frequency jumps merely decay into slope changes in the dispersion curve;

(iv) Frequency jumps or local changes in the slope of the dispersion curve occur at well-defined frequencies. Characteristic frequencies amount here to about 5.5 GHz and 4.5 GHz. They differ little from experimental values. If these frequencies are to be viewed as representative of the energy landscape under the presence of both a biasing and an applied field, then it needs being admitted that they differ little from jump frequencies calculated in the absence of any biasing field;

(v) Very unfortunately, red-shifting still proves much slower than in experiments. The second transition, especially, takes place at current densities almost twice as large as the experimental value and the last computed frequency is for the lowest damping constant also almost twice as large as the experimental frequency for $I = 10$ mA.

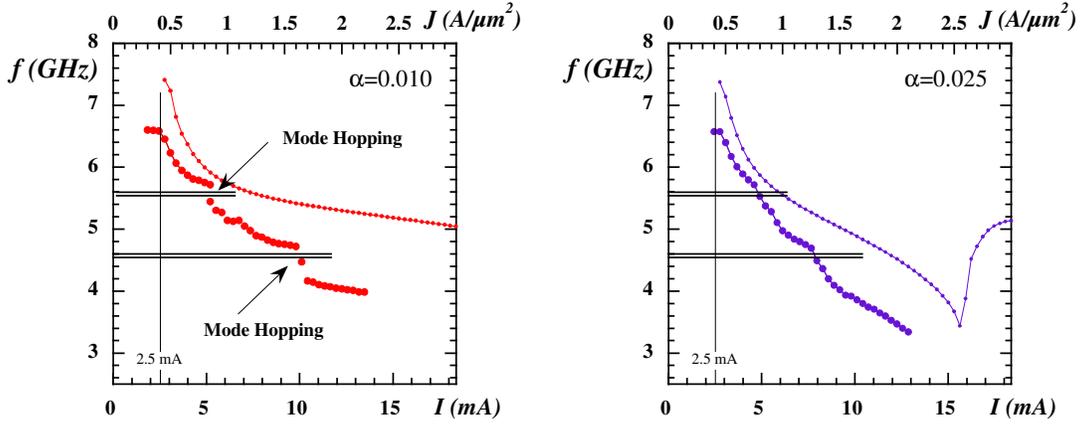

Fig. 6: Comparison of dispersion relations between macrospin and full micromagnetic simulations for two values of the damping parameter. Simulation parameters: $\chi_a = 0.5$, $M_S = 650$ kA/m, $H^{app} = 68$ mT, $H^b \approx 14.5$ mT. The effective polarization has been adjusted so that the critical current is in both cases close to 2.7 mA, i.e. $\mathcal{P}_{eff} = 37.5\%$ for $\alpha = 0.025$ and $\mathcal{P}_{eff} = 15\%$ for $\alpha = 0.010$. Temperature $T = 40$ K is independent of current.

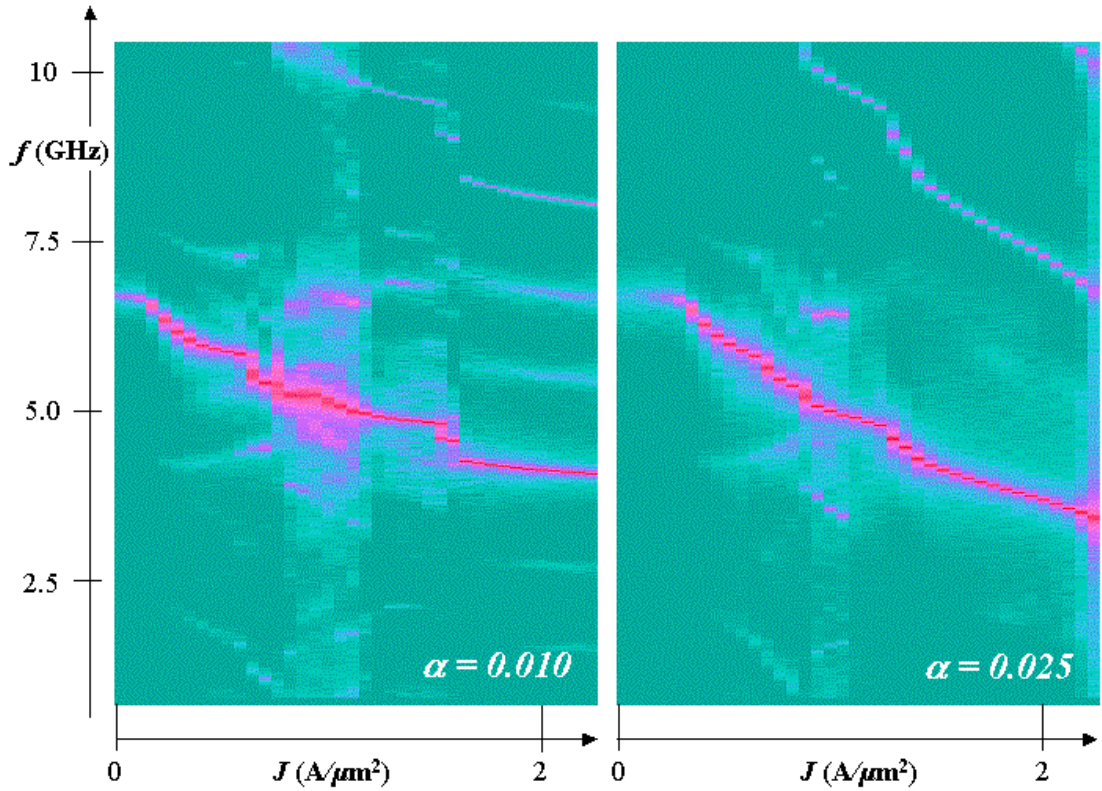

Fig. 7. Power spectral density (PSD) maps (log scale) corresponding to Fig. III.4-3. Low damping map on the left, higher damping on the right. No interpolation has been performed: the power spectral density is computed at the centre of each visible frequency segment in the images. Some current domains display clear overlapping frequencies. A non-harmonic spreading of the PSD may also be observed. Horizontal scale in agreement with current ranges in Fig. 6; vertical axis ranging from 0.67 to 10.45 GHz. Simulation parameters: $\chi_a = 0.5$, $M_S = 650$ kA/m, $H^{app} = 68$ mT, $H^b \approx 14.5$ mT. The effective polarization and temperature are the same as in Fig. 6.

On the other hand, the power spectral density remains nicely concentrated within extremely narrow line widths in these simulations as evidenced in Fig. 7, where no substantial line broadening may be observed, even for the largest current densities, except for the largest damping parameter value.

Summarizing at this point, it appears that simulations including the biasing-field do only remotely describe the characteristics of experimental dispersion curves. Best agreement is found for low damping that preserves the existence of marked frequency jumps occurring at well-defined frequencies. If the effective polarization is chosen such as to match the current threshold for the onset of precessional states, then the average red-shifting $\langle df/dJ \rangle$ proves smaller than expected by a factor close to 2. Even if taking into account some of the worse discrepancies found in either mode analysis or vortex gyration experiments quoted earlier, such a result may only appear as extremely poor.

More generally, it may be stated that

1) for a given damping constant, increasing the effective electron polarization leads to faster red-shifting. However, i) the critical current for the onset of precessional states decreases accordingly, ii) frequency jumps become gradually smeared-out,

2) for a given damping parameter and a given effective electron polarization, increasing the asymmetry parameter $\chi_a$ tends to translate the dispersion curves towards lower current densities without significant alteration of the average red-shifting velocity.

### IV.4. Nanopillar geometry: STI-dynamics features which could *not* be reproduced by simulations

Although a fair number of features of spin-torque driven magnetization dynamics could be successfully explained either by means of macrospin or full-scale micromagnetic simulations, these achievements should not be mistaken for a full understanding of this phenomenon. Several important issues must be clarified before the physical model of the spin-torque driven magnetization motion can be declared at least as reliable as standard magnetization dynamics studies based on the LLG equation. Here are some of the open problems related to experiments in the nanopillar geometry.

1) A very serious discrepancy is *the large difference between simulated [50] and measured oscillation frequencies* in the Co/Cu/Co nanostack in [49]. To start with the analysis of this problem, we first point out that the frequency is the observable that can be measured most reliably, with a very high precision, indeed. All other features of the spectral lines - line widths, peak heights, total power etc - require a much more sophisticated analysis to enable a quantitative comparison with theory, but the frequency is usually determined unambiguously. Keeping this argument in mind, the inability of simulations to reproduce experimentally measured oscillation frequency of the Co/Cu/Co nanopillar is alarming.

The first attempt to reproduce the experimental frequency was made by Lee et al [52], who used the standard Co saturation magnetization $M_S$ = 1400 G and a lower value of the Co exchange stiffness ($A_{Co}$ = 2 10$^{-6}$ erg/cm instead of the more usual value $A_{Co}$ = 3 10$^{-6}$ erg/cm). The choice for a lower exchange constant was justified by the authors [52] as a mean to take into account thermal fluctuations effects without having to solve a stochastic equation of motion (we note in passing that the validity of this tactics is presently a subject of debate, but well beyond the scope of this paper). Comparison of simulation results from [52] with measured frequencies is discouraging: in the quasichaotic regime the simulated frequencies were in the range $f_{sim} \approx$ 10 - 12 GHz (see Fig. 3d in [52]) for an applied field value equal to $H_0$ = 400 Oe. The measured frequencies are about $f_{exp} \sim$ 5.5 GHz for $H_0$ = 2000 Oe, which means that for $H_0$ = 400 Oe $f_{exp}$ would be even lower. We have simulated the quasichaotic dynamics of the Kiselev's geometry with the same magnetic parameters as in [52] and found that the quasichaotic regime starts with $f_{sim} \approx 15$ GHz for $H_0 = 2000$ Oe, i.e. nearly 3 times the measured value !

One possible reason for such a discrepancy could be non-standard values of Co magnetic parameters (first of all, the saturation magnetization $M_S$) when Co is present as a thin film. In an attempt to determine these parameters experimentally, Kiselev et al have measured the saturation magnetization of a stack consisting of several 3 nm thick extended (not patterned-processed !) Co layers and have found the value $4\pi M_S = 1 \pm 0.1\, T$, so that $M_S \approx 800$ G [49]. Our simulations [50] with this reduced $M_S$ value yielded (for $H_0 = 2000$ Oe) an oscillation frequency at the beginning of the quasichaotic regime equal to $f_{sim} \approx 11$ GHz (for the exchange stiffness $A_{Co} = 2 \cdot 10^{-6}$ erg/cm) and $f_{sim} \approx 9$ GHz (for $A_{Co} = 3 \cdot 10^{-6}$ erg/cm), still a much too high value. Apart from the unknown crystalline structure of the Co films used in [49], surface anisotropy and edge oxidation effects may play an important role in such small systems. Whether one or both of these reasons play a significant role requires careful magnetic and spatially resolved structural (and chemical) measurements performed on multilayers identical to those intended for use in spin-torque experiments. In addition, increasing the torque asymmetry can reduce the oscillation frequency in the large-angle regime (see results in [55]), but again, one needs independent experimental measurements of the corresponding asymmetry parameter to draw meaningful conclusions.

Lastly, rather unexpectedly, a better fit between micromagnetic simulations and experimental results was obtained in the red-shift regime for material parameters classical for a $Ni_{80}Fe_{20}$ alloy and a field equal to $3H_K$ [53, 39]. The frequency observed in the very low current regime could certainly not be reproduced, unless assuming that the GMR signal observed at low current densities only displays the $2f$ signature of a collinear system whereas the fundamental frequency appears at larger densities due, perhaps, to current induced magnetization oscillations within the hard layer. Such an assumption, previously unnoticed, potentially offers a credible explanation for the huge frequency drop occurring at the onset of precessional states in [49].

2) The second important problem in nanopillar experiments concerns the *dependence of the oscillation power on the current strength*. We note first, that obtaining *quantitatively* accurate values of the oscillation power (in order to enable a meaningful comparison with simulations) is a non-trivial experimental task, as can be seen from the corresponding discussion in [55]. However, this task was successfully solved for the IrMn/Py/Cu/Py nanopillar in the experiment analyzed in detail in Sec. IV.2. Comparison of experimental and numerical results (together with the frequency vs current dependence $f(I)$) is shown in Fig. 3b. The qualitative discrepancy between simulations and experiments is evident. Simulated power grows very rapidly for currents slightly above the critical value (due to the rapid growth of the oscillation amplitude associated to a nearly coherent magnetization precession), reaches its maximal value just after the first frequency jump and then slowly decreases, as a consequence of the increasing inhomogeneity of the magnetization configuration at higher currents. In contrast to this behaviour, the experimentally measured power grows much slower for currents just above the threshold and increases monotonically with current (except the dips for the current values where the frequency jumps take place).

This disagreement is especially surprising taking into account the rather good coincidence of measured and simulated frequencies (see panel (a) of the same figure). As explained above, frequency decrease with increasing current strength is a non-linear effect arising from the dependence of the oscillation amplitude on current value. The oscillation power is directly related to the amplitude, so that the large (partly qualitative !) disagreement between simulated and measured powers for the situation where the corresponding frequencies agree well, is a puzzling phenomenon, clearly requiring further investigation.

3) Turning back to simulations including the bias field arising from the pinned layer, the dispersion relation $f(J)$ is clearly unsatisfactory, but, on the other hand, the overall behaviour of the line width and power spectral density *vs* current (see Fig. 7) does not appear that far away from experimental data. In order to reach a better fit in the dispersion relation, unless admitting that the biasing field proves, for an unknown reason, much smaller than estimated from micromagnetics, is there any parameter that could still be tuned in the simulations? It seems indeed rather paradoxical that none of the simulations we know of correctly predicts the right current for the second frequency jump. Heuristically, beyond surface anisotropy and potential exchange biasing along the edges, all simulation parameters could be made current density dependent. For instance, both the damping parameter $\alpha$ and the torque asymmetry parameter $\chi_a$ could be thought of increasing with increasing current density $J$. Although highly speculative, it could also be assumed that the saturation magnetization $M_S$ decreases with increasing $J$. Test simulations seem to indicate that, at least for simulations avoiding the intrinsic deleterious effects of stair-case magnetostatics, assuming current-density dependent parameters such as $\chi_a$ or $M_S$ does not lead to line width broadening, even for the largest current densities. However, before moving ahead, theoretical insights would here be more than appreciated.

## V. Simulations of steady-state precession in the point-contact geometry

### V.1. Methodological problems

Before we start with the discussion of experiments and simulations obtained in the point contact geometry, we would like to point out that simulations of this geometry encounter several complicated methodological problems [61, 62, 48], which make this kind of simulations much more challenging than modelling of the nanopillar devices.

We remind that the corresponding experimental setup consists of a nanowire in contact with an upper layer of a multilayered ferromagnetic system. The current flows through this nanowire and magnetic layers contacted by it. If we assume for simplicity that the system is placed into an external field large enough to saturate magnetic layers, then the lower layer will preferably reflect those current electrons, which magnetic moments are oriented antiparallel to its magnetization. According to the general concept of the spin torque, such electrons create a torque acting on the upper layer magnetization, which can lead to the instability, spin-wave excitations and even switching in the upper layer, exactly as in the nanopillar geometry.

However, the point-contact geometry and hence - corresponding physics - is quite different from that of the nanopillar device. In the point contact geometry, the non-magnetic spacer thickness (5 - 10 nm) is much smaller than the contact diameter (25 - 80 nm). Thus the transversal diffusion of current electrons is negligible and reflected electrons act mainly on the magnetization in the upper layer region *under the contact*. The rest of the layer is (in a fairly good approximation) not affected by the spin torque. This means that magnetization excitations created by the spin torque in the region under the point contact, can propagate in the rest of the magnetic layer as waves emitted by a small source (but not a point source !). Hence the magnetization dynamics in this case is expected to be qualitatively different from the nanopillar case where spin waves are excited 'simultaneously' over the whole area of a magnetic nanoelement.

This difference immediately leads to several serious complications when modelling magnetization dynamics. Namely, in order to describe adequately the wave propagation and the influence of the magnetization of the rest of the layer on the point area, we need now to simulate not only the thin film region where the magnetization is excited by the spin torque (region under the contact), but also the surrounding thin film. The problem is that actual lateral sizes of a multilayer in such experiments are about ~ 10 - 20 $\mu$m, making the simulation time for the whole system prohibitively long.

The obvious solution seems to simulate a smaller area, say 1 x 1 $\mu m^2$, in the hope that this area is still much larger than the point contact size, so that our simulations will capture all the necessary physics. This hope is in principle correct, but only in principle, because the naive straightforward implementation of this idea fails due to improper boundary conditions. In a real experiment the emitted wave propagates in the laterally wide thin film until it dies due to natural energy dissipation. In simulations relying on a much smaller lateral system size, two situations are possible: (*i*) the emitted wave of a significant amplitude is reflected from the thin film border - for open boundary conditions, or (*ii*) the wave coming from the system replica(s) - for periodic boundary conditions - penetrates into the simulated area. In both cases we deal with an artificial 'incoming' wave, which propagates from the borders of the simulated region towards the point contact area. The interference of this second wave with the magnetization oscillations within the point contact region may lead to unpredictable and physically completely meaningless results, especially taking into account that this interference is particularly strong - both waves have the same frequency.

The 'clean' solution of this problem requires the analytical derivation of perfectly absorbing boundary conditions which should be imposed on the magnetization at the system borders in order to completely absorb the incident wave. However, this derivation has not been done yet and it is not clear whether it can be done at all. For this reason we have proposed a numerical trick basing on the introduction of a space-dependent dissipation [61, 62]. Namely, dissipation constant at and near the point contact area is set to real physical dissipation value, so that oscillations of the point contact area and the emitted wave are not affected. With the growing distance to the contact center the dissipation increases, so that near the simulation area borders it is so large that the wave passing through this area loses all its energy and 'disappears'. The spatial profile of this dissipation increase and the maximum dissipation value at the area border should be chosen carefully: if the profile is too steep, wave reflection can still occur due to too rapid a change of the medium properties. The rigorous theory how to choose such a profile is also not available (although numerical criteria whether the profile is chosen correctly, do exist [62, 63]), so the corresponding operation is still more art than science.

### V.2. Various dynamic modes and their excitation threshold

Experimental results on point contact systems**,** which can be used for a meaningful comparison with simulations, are very rare [43, 44, 64, 65], what is surely one of the major reasons why the corresponding dynamics is understood much poorer that for nanopillar devices.

After the methodical difficulties mentioned in the previous section were resolved and the first rigorous simulations of STI dynamics in point contact devices could be performed, it has turned out that for this kind of experiments not even a qualitative agreement between simulations [66] and 'real' data [43] could be claimed. Two major qualitative discrepancies between theory and experiments were: (*i*) for the current just after the oscillation onset, the simulated frequency was almost two times larger than measured experimentally and (*ii*) even when the Oersted field of the electric current was neglected, two steady-state precession regimes were found in simulations (with nearly opposite directions of the precession axes, very different oscillation frequencies and different frequency vs. current characteristics), when only one regime was observed in the experiment.

The discrepancy between the oscillation frequencies should indeed be considered as a qualitative, and not a quantitative one. Namely, the frequency measured experimentally $f_{exp} \approx 7.6$ GHz (see Fig. 1 in [43]) was not only nearly two times smaller than the value $f_{sim} \approx 12 - 13$ GHz found numerically; the major problem was that the experimental frequency was *below* the homogeneous FMR frequency $f_{FMR} = 8.4$ GHz for the system studied in [43], i.e., a thin film made of Py with the saturation magnetization $M_S = 640$ G, placed in an external field $H_0 = 1000$ Oe. This means that the mode detected experimentally could not be the propagating wave, as it was the case in numerical simulations for the regime observed directly after the oscillation onset. In this first numerical study [66] a regime with magnetization oscillations *localized* under the point contact area was also found – this was the 2$^{nd}$ regime mentioned above. However, the nature of this localized oscillation mode was not clarified in [66].

Detailed theoretical studies of magnetization oscillations induced by a point-contact spin torque were performed by Slavin et al. [47, 67] using analytical methods suitable for non-linear dynamics. In [47, 67] it was shown that there exist at least two solutions for the equations of the magnetization motion under the influence of the Slonczewski torque. The first solution which was already found by Slonczewski [20] is the standard solution of the linearized equation of motion and represents the propagating wave with the wave vector $k \sim 1/R_c$ ($R_c$ being the radius of the point contact). The frequency $f(k)$ for this regime is much higher than the FMR frequency, because the corresponding wave vector is relatively large. As a solution of the linearized equation, it could have an arbitrarily small amplitude. Oscillations in the first regime – just after the critical current – found in [66] do correspond to this propagating solution.

Slavin et al. [67] have found out, that the 'native' (non-linearized) equations of motion in the point-contact geometry possess also another solution – the so called non-linear 'bullet' mode. Magnetization oscillations in this mode are localized within the point contact area, i.e., their amplitude $b_0$ decreases exponentially with distance $r$ from the point contact center ($b_0 \sim \exp(-r/R_{dec})$, decay radius $R_{dec} \sim R_c$); we recall that $b_0 \sim 1/r^{1/2}$ for the linear mode in 2D. The frequency of such a localized mode is smaller than the lowest possible frequency of the propagating mode $f_{FMR}$; this is an inherent feature of a localized solution. The amplitude of this mode cannot be arbitrarily small: as often the case for non-linear oscillations, the amplitude acquires a finite value immediately above the critical current threshold.

Another interesting feature of this mode is also due to its localization. Namely, being localized, the 'bullet' does not lose energy due to energy radiation in form of a propagating wave and thus can have a smaller excitation threshold (smaller critical current required for its excitation), despite the fact that right after the excitation it should have a significant amplitude. This analytical prediction was the subject of a controversial debate. Namely, the localized mode reported in [66] and described in more detail in [61] had many common features with the 'bullet' described by Slavin et al.: spatial localization near the point contact area, frequency lower than $f_{FMR}$, nearly homogeneous structure of the mode kernel. On the other hand, in simulations the localized mode was excited *after* the linear (Slonczewski) mode, i.e., for currents larger than the linear mode threshold, so that the question about which mode was observed in a real experiment, still remained unclear.

This controversy was resolved in the work of Consolo et al. [63], who could show that in order to observe *numerically* the non-linear bullet with the excitation threshold smaller than that for the linear mode, one should not increase the current, but rather move from large currents - where the 'bullet' does already exist - to smaller ones. In this case the non-linear solution persisted down to currents smaller than the linear mode threshold, confirming the prediction from [67]. Stated otherwise, magnetization excitation dependence on current strength for the point contact geometry demonstrates a hysteretic behaviour at $T = 0$.

Essentially the same explanation for the discrepancy discussed above was suggested in [48], where the *energy* behaviour for corresponding modes was studied as a function of *increasing* current. It was found, that although the non-linear mode was excited later than the linear one with increasing current, the average system energy for this non-linear mode was significantly lower than for the Slonczewski solution (propagating wave), so that the Slonczewski mode was only metastable. It was excited first because simulations done for $T = 0$ could not provide fluctuations required to excite a non-linear mode with a finite excitation energy threshold. So at the present state of knowledge the most likely explanation of the results described in [43] is that due to thermal fluctuations the non-linear 'bullet' mode was observed there. Experiments from [43] were performed at room temperature, so that this mode could be excited although observations were made under increasing current.

As a matter of facts, it quickly turned out that from the theoretical point of view the situation proves even more complicated. Already in [62] another non-linear localized mode was detected. In contrast to the solution found in [67], this additional mode had a kernel with a highly complicated magnetization structure consisting of two vortex-antivortex pairs (see Fig. 8d). More detailed studies have shown [48] that for currents where both non-linear modes ('bullet' and vortex-antivortex) can exist, the energy of the vortex mode in significantly lower, so that this mode should be actually observed. Moreover, we have also shown, that some perturbations present in real system (e.g., the dipolar field from the 'fixed' layer) strongly favour the transition from the 'bullet' to the vortex mode. The reason why Rippard et al. have found the 'bullet' mode, thus remains unclear.

Experimentally this second (vortex-antivortex) mode type can be distinguished from the 'bullet'-like oscillations using the following criteria.

First, the frequency of the vortex mode is much lower than that of the 'bullet' mode; e.g., for conditions corresponding to the experiment of Rippard et al. ($M_S = 640$ G, $H_0 = 1000$ Oe) [43] we have found $f_{bull} \approx 2 \cdot f_{vort}$ [48]. In smaller external fields the frequency of this mode can be as low as ~ 200 – 300 MHz. The reason for such a low frequency is that the oscillations of the average magnetization in this mode occur due to the creation-annihilation of vortex-antivortex pairs, which is a rather slow process. As mentioned above, frequency of the microwave output signal can be measured experimentally and determined in simulations very reliably, so that this criterion is relatively easy to apply, if the real system is characterized carefully enough.

Second, the total oscillation power in the vortex mode (averaged over the point contact area) is significantly smaller – up to several times – than that of the 'bullet' mode. This is due to a more complicated magnetization configuration of the vortex mode kernel. Usage of this criterion requires reliable quantitative measurements of the microwave oscillation power produced by the point contact device.

Third, although the frequency of both localized modes is lower than $f_{FMR}$, so that they can not emit power in form of 'normal' propagating waves, they still radiate energy in form of spatially localized solitons. This radiation is highly anisotropic and the anisotropy patterns are very different for 'bullet' and vortex modes. However, to apply this criterion, one needs to perform measurements of the magnetization oscillations for frequencies in the GHz range and with the spatial resolution ~ 50 nm, which is extremely difficult. On the other hand, any attempt to obtain the synchronized point-contact oscillators (keyword 'frequency locking') [46, 45] in the field-in-plane geometry requires detailed studies of the energy radiated from point contacts, so that further progress on this topic is quasi unavoidable.

Concluding this subsection, we would like to mention, that our preliminary simulation results indicate that the vortex-antivortex mode could be responsible for the low-frequency oscillations reported very recently in [68], where experimental data for the point-contact geometry with a somewhat higher contact radius ($R \approx 40 - 50$ nm) are discussed.

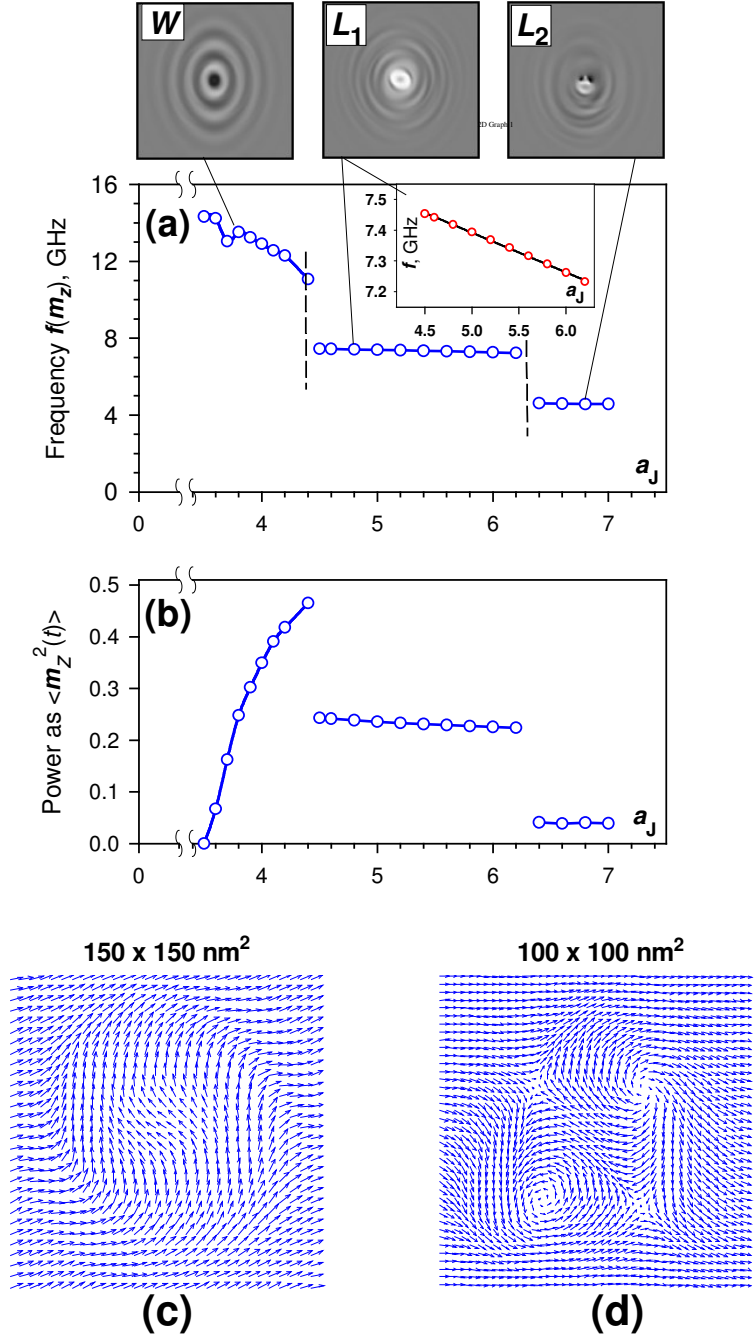

Fig. 8. Dependencies of the oscillation frequency $f$ (panel (a)) and power $P$ (panel (b)) on the spin torque magnitude $a_J$ (proportional to the current strength $I$) for the point contact setup. Transitions from the *extended* wave mode $W$ to the *localized* mode $L_1$ and from $L_1$ to the 2$^{nd}$ *localized* mode $L_2$ are clearly seen. Snapshots of the in-plane magnetization projection perpendicular to the applied field direction for all mode types are shown as grey-scale maps. Panels (c) and (d) display the in-plane magnetization arrow images of the localized mode cores: (c) - $L_1$ mode (every second moment is shown), (d) - $L_2$ mode.

# VI. Importance of a sample characterization

It was known from the very beginning of research on spin-torque induced magnetization dynamics that one would need extremely small sizes of the corresponding structural units (devices) in order to separate the STI-induced phenomena from effects arising from the Oersted field generated by the large electric current required to induce steady-state precession or even switching of the sample magnetization. Indeed, most of the high-quality experiments used to compare theoretical predictions with experimental findings were performed on very small nanopillars (lateral sizes ~ 50 - 200 nm) or point contact devices with contact diameter 25 - 80 nm.

Characterization of magnetic structures with such small sizes presents a new challenge to the experimentalists: one should not only know the exact magnetic parameters of the extended films which are used to produce, e.g., nanopillar devices, but also be able to measure structural, chemical and magnetic properties of the nanoelements themselves.

First such a knowledge is indeed crucial if a real theoretical understanding of spin-torque physics is sought for, and not only a qualitative description making use of analytical perturbation theories or the macrospin approximation. Micromagnetics is credited with a high predictive power, because for 'standard' (not involving spin-torque) magnetization dynamics is it able to describe and explain experimental data with a high precision. On top of examples quoted in the introduction, a good recent example can be found, e.g., in [69], where non-trivial discrete spectra of magnetization oscillations in rectangular nanoelements (measured via the Brillouin light scattering) were reproduced with a very satisfactory quantitative agreement.

Hence, if aiming at such a quantitative agreement for the STI dynamics also, we actually need a precise characterization of each particular device the dynamics of which needs to be modelled. The reason why we really need characteristics of a particular sample is the large sample-to-sample variation of experimental data. For example, looking at the *f(I)* curves in Fig. 1 from [49] and comparing them with the phase diagram of the oscillation power in (*I*-$H_0$)-coordinates (Fig. 2 ibid.), we can immediately recognize that the two devices which data are shown on these two figures, produce significantly different dynamical output. The same conclusion naturally results from the wide region of fitting parameters which, according to [49], was required to fit the data of various samples. In [55] frequency vs current curves are explicitly shown (see Fig. 6 in [55]) and clearly demonstrate significant variations from sample to sample.

The most important features of the nanodevices (leaving aside evidently required characteristics like the saturation magnetization, exchange constant, average bulk anisotropy etc) which one definitely needs for a quantitative - and often even for a qualitative - data analysis using micromagnetics, are the following:

1) *Polycrystalline structure* of the magnetic films used to produce nanopillars or point contact devices. Here we need the average grain size, grain texture (if present), and the type of crystallites (crystal lattice). These data are especially important when analyzing data from devices based on Co or CoFe films, because the magnetocrystalline anisotropy of these materials can be high and there are even two possible crystallographic phases for Co - *hcp* and *fcc*. In our detailed analysis we have shown [50] that the magnetization dynamics of Co nanoellipses can qualitatively depend on the type of Co crystal grains.

2) *Surface and edge anisotropy* are also very important for the devices under discussion, because, due to the very small element sizes, surface and edge effects play a very important role. We strongly suspect that the large sample-to-sample variations of the observed dynamics found, e.g., in [49] and [55] (and many other papers) is at least partly due to an imperfect element shape and edge anisotropy.

3) *Irregularities of the chemical composition*. In particular, it is known that for some specific patterning techniques nanostructure edges are oxidized to some extent. It is also known, that oxidation may affect not only the saturation magnetization, but also the damping constant of a magnetic material, which is also a very important parameter when performing simulations.

4) In exchange biased spin valve systems the *orientation of the exchange bias direction* is an essential information. It is hardly possible to change the exchange bias orientation directly, so one should perform the measurements of quasistatic hysteresis loops on such spin valves using, e.g., the GMR effect and then try to obtain the required direction by fitting this loop.

## VII. Conclusion

Keeping in mind that this series of 'Current Perspectives' papers should present the very recent state of the art and simultaneously 'reflect the personal point of view' of the authors, it is almost impossible to write a conclusion in the sense known from 'standard' scientific contributions. For this reason we would like to confine ourselves to the following final remarks.

First, nowadays everybody is used to numerical simulation analysis as an extremely useful tool for interpreting experimental results and gaining new insights into the physics of various processes. On the other hand, for micromagnetics in particular, this acceptance of numerical simulations coexists (in a very strange way) with the point of view, that when full-scale simulations give a *poorer* agreement with experiment than oversimplified models (first of all the macrospin approximation), then one should abandon any attempt to obtain a good agreement using full-scale micromagnetics and return to these simple models. So, especially in connection to micromagnetic simulations of the STI dynamics, which still demonstrate many important discrepancies with experimental results, we would like to emphasize the following. Full-scale simulations are supposed, first, to take into account *more* (and *not* fewer) known features of the studied systems than simplified (semi-)analytical models, and second, to use *fewer* (if any) adjustable parameters. This means, that if full-scale simulations demonstrate a *poorer* agreement with experiment than the macrospin analytics and/or simulations, then we should realize that we *do not understand* some crucial physical properties of the system under study. Hence the disagreement between simulations and experiment obviously calls for *further research*, and *not* for returning to oversimplified models where a good fit to experimental data can still be achieved by adjusting several parameters without having a real physical justification for such an adjustment.

Second, meaningful full-scale simulations require an accurate knowledge of the underlying system parameters, obtained from independent sources (preferably high-quality experiments). Being still very time-consuming, full-scale simulations usually do not allow for an exploration of the whole parameter space, so that the corresponding experimental input is really mandatory. Applied to simulations of the STI-dynamics, this statement means, of course, that we need *independent* analytical calculations and experimental measurements of the STI-specific characteristics like the current polarization degree for various FM/NM interfaces. However, probably even more important is the understanding, that we need not only the *average* values of the saturation magnetization, exchange stiffness etc. of nanomagnets under study, but also the information how the processing of corresponding devices may change magnetic parameters *locally*, especially near those edges of the nano-structures which are influenced (or even created) by this processing. As shown above, such information is crucially important for the modelling of nanodevices due to their extremely small sizes, resulting in the increasing influence of the structure imperfections and local parameter variations.

And last, but not least, recent simulations have shown that the influence of the Oersted field may have a *qualitative* impact on the observed dynamics. This insight raises the question how to correctly calculate this field in the state-of-the-art spin transfer devices. Being in principle straightforward and purely technical, such calculations still require an exact knowledge of the 3D distribution of an electric current in such systems, a task which is definitely out of the scope of the micromagnetics and requires a significant separate effort for its solution.

**Acknowledgements:** JM gratefully acknowledges support from the European Community programme "Structuring the ERA", under contract MRTN-CT-2006-035327 SPINSWITCH, DB thanks the German Research Society for its support in frames of the SPP "Ultrafast magnetization dynamics" (project BE 2464/4).